\documentclass{article}

\usepackage[preprint]{neurips_2023}
\usepackage{caption}

\usepackage[utf8]{inputenc} 
\usepackage[T1]{fontenc}    
\usepackage{hyperref}       
\usepackage{url}            
\usepackage{booktabs}       
\usepackage{amsfonts}       
\usepackage{nicefrac}       
\usepackage{microtype}      
\usepackage{xcolor}         
\usepackage{amsmath}
\usepackage{amssymb}
\usepackage{mathtools}
\usepackage{amsthm}
\usepackage{bbm}
\usepackage{multirow}
\usepackage{caption}
\usepackage{subcaption}

\usepackage[capitalize,noabbrev]{cleveref}

\theoremstyle{plain}

\theoremstyle{definition}

\theoremstyle{remark}

\usepackage[textsize=tiny]{todonotes}

\title{A Distribution-free Mixed-Integer Optimization Approach to
          Hierarchical Modelling of Clustered and Longitudinal Data}

\author{
   Madhav~Sankaranarayanan\\
   Department of Biostatistics\\
   Harvard T.H.Chan School of Public Health\\
   Boston, MA 02115 \\
   \texttt{madhav\_sankaranarayanan@g.harvard.edu} \\
  \And
   Intekhab~Hossain \\
   Department of Biostatistics \\
   Harvard T.H.Chan School of Public Health\\
   Boston, MA 02115 \\
   \texttt{ihossain@g.harvard.edu} \\
   \And
   Tom~Chen\\
   Department of Population Medicine\\
   Harvard Medical School/Harvard Pilgrim Health Care Institute\\
   Boston, MA 02115 \\
   \texttt{tchen@hsph.harvard.edu} \\
}

\begin{document}

\maketitle

\begin{abstract}

Recent advancements in Mixed Integer Optimization (MIO) algorithms, paired with hardware enhancements, have led to significant speedups in resolving MIO problems. These strategies have been utilized for optimal subset selection, specifically for choosing $k$ features out of $p$ in linear regression given $n$ observations. In this paper, we broaden this method to facilitate cluster-aware regression, where selection aims to choose $\lambda$ out of $K$ clusters in a linear mixed effects (LMM) model with $n_k$ observations for each cluster. Through comprehensive testing on a multitude of synthetic and real datasets, we exhibit that our method efficiently solves problems within minutes. Through numerical experiments, we also show that the MIO approach outperforms both Gaussian- and Laplace-distributed LMMs in terms of generating sparse solutions with high predictive power. Traditional LMMs typically assume that clustering effects are independent of individual features. However, we introduce an innovative algorithm that evaluates cluster effects for new data points, thereby increasing the robustness and precision of this model. The inferential and predictive efficacy of this approach is further illustrated through its application in student scoring and protein expression.

\end{abstract}

\section{Introduction}
\label{submission}

\subsection{Problem setup}
Linear mixed-effects models (LMMs) constitute a significant toolset for analyzing hierarchical data. These models find extensive applications in various domains such as healthcare, social sciences, and longitudinal studies, among others \citep*{Liao2016-ps, Balasankar_2021, Shiratori2020-sl}. The conventional approach to account for clustering variables in an ordinary least squares (OLS) context becomes computationally impractical as the number of clusters scales up, requiring a separate parameter for each cluster. \citet{Laird_Ware_2023} introduced the LMM formulation that conceptually treats each cluster's characteristics as random variables:
\begin{align}
\label{eq:LMM}
Y_{ki}\mid \mathbf{X}_{ki}, \mathbf{Z}_{ki}, \boldsymbol{\gamma}_k \sim \mathcal{N}(\boldsymbol{X}_{ki}^\intercal\boldsymbol{\beta} + \mathbf{Z}_{ki}^\intercal\boldsymbol{\gamma}_k, \sigma_\varepsilon^2), \qquad \gamma_k \sim F(\theta_\gamma)
\end{align}
This equation defines $Y_{ki}$ as the $i$th observed outcome of cluster $k$, $\boldsymbol{X}_{ki} \in \mathbb{R}^{P+1}$ as the corresponding covariates (including intercept), and $\sigma_{\varepsilon}^2$ as the observation-specific noise. The variable $\boldsymbol{\gamma}_k$ represents the random effects, implying a clustering interaction effect between cluster $k$ and covariates $\mathbf{Z}_{ki} \in \mathbb{R}^{Q+1}$ (including a pure cluster effect from a \textit{random intercept}). The random effects are assumed to follow a distribution $F$ parameterized by $\theta_\gamma$. Overlooking random effects equates to neglecting clustering effects, which can lead to flawed inference \citep{Ntani_Inskip_Osmond_Coggon_2021}. While the LMM formulation provides a convenient and efficient approach, especially when $\dim(\theta_\gamma) \ll K$, the fundamental assumptions of Eq (\ref{eq:LMM}) introduce limitations in terms of robustness and generalizability.

Classical LMMs often assume normal distribution for $\boldsymbol{\gamma}_k$, which can be overly restrictive, especially when dealing with missing data or outliers. This paper's focus lies not on consistent predictions of random effects $\boldsymbol{\gamma}_k$, but rather on sparsistent predictions that distinguish the cluster effects significantly deviating from the rest, assigning the remaining cluster effects to zero. To this end, we focus on the following formulation:
\begin{align}
\label{eq:LMM-MIO}
\min_{\boldsymbol{\beta}, \boldsymbol{\Gamma}} \sum_{k=1}^{K}\frac{1}{2}\|\mathbf{Y}_k - \mathbf{X}_k \boldsymbol{\beta} - \mathbf{Z}_k \boldsymbol{\gamma}_k\|_2^2 \quad \text{subject to} \quad \|\boldsymbol{\gamma}_{r}'\|_0 \le \lambda_r \quad \text{for } r = 0, \cdots, Q
\end{align}
where $\mathbf{Y}_k \in \mathbb{R}^{n_k}$, $\mathbf{X}_k \in \mathbb{R}^{n_k \times (P+1)}$,  $\mathbf{Z}_k \in \mathbb{R}^{n_k \times (Q+1)}$ are row-stacked matrices of $Y_{ki}, \mathbf{X}_{ki}, \mathbf{Z}_{ki}$ for each cluster $k$,
$\boldsymbol{\gamma}_{r}' \in \mathbb{R}^{K}$ are the random effects over all $K$ clusters associated with covariate $r$ in $\mathbf{Z}_k$, and $\ell_0$-norm is defined as $\|\boldsymbol{x}\|_0 = \sum_{i=1}^{n}\mathbb{I}(x_i \neq 0)$. As we move forward, it's crucial to recognize that $\lambda_r$ possesses an immediate interpretation as a \textit{sparsity} parameter, directly controlling the number of non-zero elements in $\boldsymbol{\gamma}_{r}'$.

The crux of this paper lies in addressing problem \ref{eq:LMM-MIO} using contemporary optimization methods, specifically mixed integer optimization (MIO), and a discrete extension of first-order continuous optimization methods. Through an array of synthetic and real datasets, we illustrate our approach's capability in solving problems where $N$ lies in the 1000s and $K$ in the 100s within a few minutes. As far as we are aware, this is the first instance of applying MIO to the identification and selection of significant random effects, a novelty that has led us to term the problem as \textit{significant cluster selection} (SCS). We use the term tractability here not to imply the usual polynomial solvability for problems, but rather to indicate the ability to solve problems of realistic size with associated certificates of optimality, within timeframes appropriate for the applications under consideration.

\subsection{Brief context and background}
A vast array of studies has concentrated on enhancing the robustness of linear mixed models \citep{lin2008estimation,lin2008longitudinal,pinheiro2001efficient,verbeke1996linear, chen2020inference}. These studies primarily aim to generalize the distribution of $\boldsymbol{\gamma}_k$ to facilitate better consistent predictions, rather than sparsistent predictions. In a non-clustered setting, numerous works have focused on the best subset for $\beta$ coefficients, including AIC approaches \citep{zhang2016variable}, Lasso \citep{zou2006adaptive}, and $\ell_0$ MIO \cite{Bertsimas_King_Mazumder_2015}. However, research focusing on SCS is virtually non-existent; \cite{geraci2020family} formulated Laplace-distributed random effects, which equivalently performs Lasso on the random effects, but the authors did not explore its use for SCS further.

Although a Lasso-like approach could be considered for the random effects $\boldsymbol{\gamma}_k$, and despite its favorable statistical properties, Lasso has been found to have several shortcomings in the best subset selection problem for $\boldsymbol{\beta}$ coefficients. In the presence of noise and correlated variables, Lasso tends to include a large number of non-zero coefficients (all of which are shrunk towards zero), including noise variables, to deliver a model with good predictive accuracy. This leads to biased regression coefficient estimates, as the $\ell_1$-norm penalizes larger coefficients more severely than smaller ones. Best subset selection procedures, in contrast, include a variable in the model without any shrinkage, thereby draining the effect of its correlated surrogates. As regularization increases, Lasso sets more coefficients to zero, but in the process, it may exclude true predictors from the active set. Therefore, once certain sufficient regularity conditions on the data are violated, Lasso becomes suboptimal as a variable selector and in terms of delivering a model with good predictive performance. This gap between the capabilities of best subset selection and Lasso is well-documented in the statistics literature, supported by both empirical and theoretical evidence \citep{greenshtein2006best, mazumder2011sparsenet, raskutti2011minimax, zhang2014lower}.

\subsection{Our approach}

In this work, we present a novel formulation of the mixed-effects regression model, which is both distribution-free and interpretable. Notably, our model abandons the independence assumption, and instead, utilizes the estimated cluster effects to enhance the predictions for new data points. The primary contributions of this paper are as follows:

\begin{itemize}
\item \textbf{MIO reformulation}: We leverage the mixed-integer optimization (MIO) framework to reformulate the mixed-effects regression model into an equivalent MIO problem, thereby circumventing the need for distributional assumptions. Our framework effectively accounts for the intrinsic correlations within clustered data (Section \ref{all_methods}).
\item \textbf{Enhanced interpretability and valid inference}: By addressing the SCS problem, our method lends an interpretability to the cluster effects, which can be categorized in a systematic manner. Our model furnishes estimates of both $\beta$ and $\gamma_k$, thus facilitating the identification of both population-level ($\beta$) and cluster-specific ($\gamma_k$) effects.
\item \textbf{Informed cluster predictions}: As our model does not require the assumption of independence between cluster effects and the clusters themselves, we can enhance our predictions for data points from new clusters (e.g., a new patient from a hospital not included in the training data). We achieve this by augmenting our MIO framework with a classification tree that maps from features ($X_k$) to predicted cluster effects ($\widehat{\gamma_k}$).

\end{itemize}

We empirically validate the efficacy of our approach through comprehensive simulation studies, in both predictive and inferential capacities (Section \ref{simu_results}). Further, we apply our methodology to real-world datasets, involving student performance analysis and protein expression (Section \ref{data_example}).   
\vspace{-2mm}
\section{Methods}\label{all_methods}

\subsection{ General MIO formulation}

We will now detail the mathematical formulation of this optimization problem given in \ref{eq:LMM-MIO}
We introduce auxiliary variable $\boldsymbol{A}\in\{0,1\}^{n 
\times k}$ such that $\boldsymbol{A}_{ij} = 1$ if individual $i$ belongs to cluster $j$. Let $\overline{\mathbf{Y}} = \left[\mathbf{Y}_1 : \cdots : \mathbf{Y}_K\right]$ be the concatenated vector-stack of all outcome vectors $\mathbf{Y}_k$ with length $n:=\sum_{k=1}^{K} n_k$, and similarly defined for $\overline{\mathbf{X}}$. Then, we may rewrite Eq \ref{eq:LMM-MIO} as
\begin{align}
\min_{\boldsymbol{\beta},\boldsymbol{\Gamma}} \sum_{i=1}^{n}\Big(\overline{\mathbf{Y}}_{i}-\boldsymbol{\beta}^\intercal\overline{\mathbf{X}}_{i} - \sum_{r=1}^{Q}((\boldsymbol{z}_r\mathbf{1}_K^\intercal)\circ \boldsymbol{A})\boldsymbol{\gamma}_{r}' \Big)^2 \quad \text{subject to} \quad ||\boldsymbol{\gamma}_{r}'||_0\leq \lambda_r \text{ for }r = 0, \cdots, Q
\end{align}
where $\circ$ is the Hadamard product, $\mathbf{1}_K$ is a length-$K$ vector of 1's, and $z_0 = \mathbf{1}_K$ represents the random intercept. Upon a further augmentation $\boldsymbol{\widetilde{X}} = \left[\boldsymbol{\overline{X}}:\left(\boldsymbol{Z}\otimes\mathbf{1}_K^\intercal\right)\circ\left(\mathbf{1}_{1+Q}^\intercal\otimes \boldsymbol{A}\right)\right]$ and $\boldsymbol{\widetilde{\beta}} = \left[\boldsymbol{\beta}^\intercal:\boldsymbol{\gamma_0}^\intercal:\boldsymbol{\gamma_1}^\intercal:\cdots:\boldsymbol{\gamma_Q}^\intercal\right]^\intercal$, where $\otimes$ denotes the Kronecker product, we have the succinct representation
\begin{align} \label{eq:mio-succinct}
\min_{\boldsymbol{\beta},\boldsymbol{\Gamma}} \sum_{i=1}^{n}\Big(\mathbf{\overline{Y}}_{i}-\boldsymbol{\widetilde{X}}_{i}\boldsymbol{\widetilde{\beta}} \Big)^2 \quad \text{subject to} \quad ||\boldsymbol{\gamma}_{r}'||_0\leq \lambda_r \text{ for }r = 0, \cdots, Q
\end{align}
which can be solved akin to a best subset selection problem. Specifically, in order to induce the $\ell_0$-constraint in this framework, a ridge penalty term ($\ell_2$) is added to allow for convergence \citep{Bertsimas_King_Mazumder_2015}:
\begin{align} \label{eq:mio-ridge-addition}
\min_{\boldsymbol{\beta},\boldsymbol{\Gamma}} \sum_{i=1}^{n}\Big(\mathbf{\overline{Y}}_{i}-\boldsymbol{\widetilde{X}}_{i}\boldsymbol{\widetilde{\beta}} \Big)^2 + \mu\|\boldsymbol{\widetilde{\beta}}\|_2^2 \quad \text{subject to} \quad ||\boldsymbol{\gamma}_{r}'||_0\leq \lambda_r \text{ for }r = 0, \cdots, Q
\end{align}
from which can implement the core ``outer'' approximation algorithm. We will demonstrate this formulation in the case of random intercepts ($Q = 0$), where we specialize to  $\boldsymbol{\widetilde{X}}=[\boldsymbol{X}:\boldsymbol{A}]$, and extended parameter vector $\boldsymbol{\widetilde{\beta}}=[\boldsymbol{\beta}^\intercal:\boldsymbol{\gamma_0}^\intercal]^\intercal$. Sparsity induction can be done using indicator variables  $\widetilde{s}_r := \mathbb{I}(\widetilde{\beta}_r\neq0)$, $r = 1, \cdots, p+1+K$. Since our interests lie within the $\boldsymbol{\gamma}_0$ portion of $\boldsymbol{\beta}$, we set $\widetilde{s}_r = 1$ for $1 \le r \le p+1$ and concentrate decision variables $\{\widetilde{s}_r\}_{r=p+2}^{p+1+K}$ in solving our ``inner''  optimization problem:
\begin{align}
\min_{\widetilde{S}} c(\widetilde{S}) := Y^\intercal \left(I-\boldsymbol{\widetilde{X}}\widetilde{S}\left(\mu I+\widetilde{S}\boldsymbol{\widetilde{X}}^\intercal\boldsymbol{\widetilde{X}}\widetilde{S}\right)^{-1}\widetilde{S}\boldsymbol{\widetilde{X}}^\intercal\right)Y \quad \text{subject to} \quad \sum_{i=p+2}^{p+1+K}\widetilde{s}_r\leq\lambda_1
\end{align}
where $\widetilde{S} = \text{diag}(\widetilde{s})$. We implement the outer approximation algorithm, using cutting planes \citep{outer}, on the set of $\widetilde{S}$, and we recover our original vector $\widetilde{\beta}$. Additionally, the optimality of our solution and the rate of convergence to said solution has been investigated in \citep{conv_outer} and \cite{conv_mio}. We provide a theoretical justification for this methodology in the Supplemental Material.

\vspace{-2mm}
\subsection{Algorithmic pipeline}
\begin{figure}[h]
    \centering
    \includegraphics[width = 0.75\textwidth]{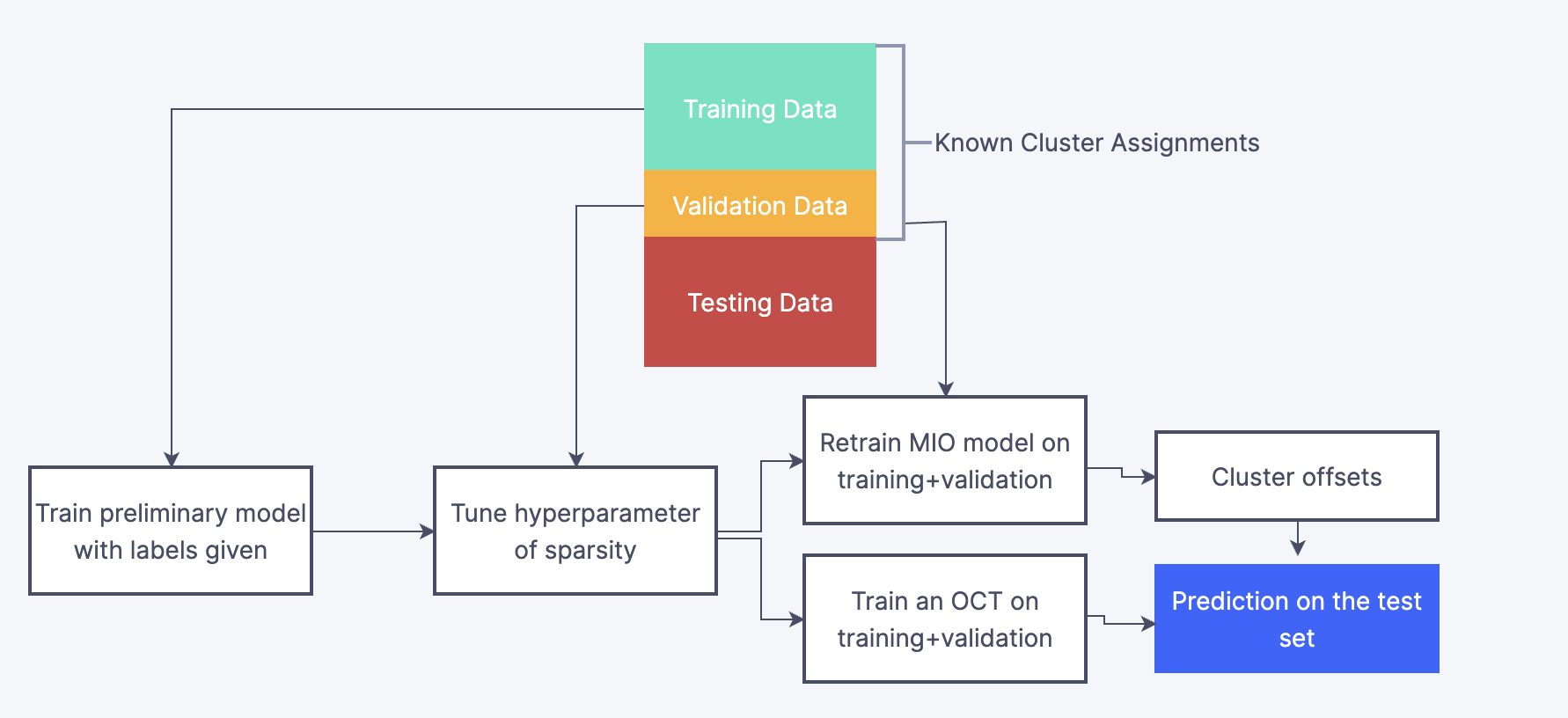}
    \caption{Flowchart of the experimental pipeline, also outlining the key components of our model}
    \label{fig:pipeline}
\end{figure}

\vspace{-1mm}
We will explain the algorithmic pipeline of our implementation. The implementation of this code was done in Julia, using Gurobi \citep{gurobi} as the optimization tool. Figure \ref{fig:pipeline} shows a flowchart for the algorithm.

\subsubsection{Modelling}

Initially, we have our training data, and the specific cluster labels of the training observations. In order to run the MIO solver on this dataset to retrieve $\beta$, we must first ascertain the choice of sparsity. We perform a grid search for the sparsity level, using MSE on a labelled validation set as a metric. Using the chosen level of sparsity, the model is refit on the augmented training set (combination of the training and validation set). The choice of the remaining hyperparameters in the MIO formulation is arbitrary. For example, the hyperparameter for the $\ell_2$ penalization can take any relatively small value, as it does not affect the accuracy or computation time of the MIO algorithm. Using Gurobi to solve the outer approximation problem, we obtain $\beta$ and $\gamma$, that is the regression coefficients and the cluster-based intercepts, an artifact of the optimization setup. 

\subsubsection{Classification and assignments}

To extrapolate cluster information from the training dataset, we propose a method that combines cluster effect data and a trained classifier to enhance predictions. A central objective is to make precise predictions for new observations that may not belong to any of the preexisting clusters.

The most straightforward prediction approach would involve the use of only the population-level estimates. In other words, for a new observation $\boldsymbol{X}_\nu$, we would predict $Y_\nu = \mathbf{X}_\nu \boldsymbol{\beta}$. We opt for the implementation of ``interpretable'' classification algorithms, such as CARTs \citep{breiman2017classification} or OCTs \citep{Bertsimas_Dunn_2017}, which facilitate the interpretation of underlying clustering. The classifier is executed on the test set, and class distributions are procured. Given a new observation $\boldsymbol{X}_\nu$ with a classification distribution of $\pi_\nu$, we propose two potential usage scenarios:

\begin{itemize}
\item \textbf{Hard Assignment}: Select the most probable cluster and directly incorporate the cluster effect into the fitted values: $Y_\nu = \boldsymbol{X}_\nu\boldsymbol{\beta}_{MIO}+\gamma_{\mathrm{argmax}({\pi_\nu})}$
\item \textbf{Soft Assignment}: Compute a weighted average (using $\pi_\nu$ as weights) of the cluster effects and incorporate the result into the fitted values: $Y_\nu = \boldsymbol{X}_\nu\boldsymbol{\beta}_{MIO}+\pi_\nu^\intercal\boldsymbol{\gamma}$
\end{itemize}

It is noteworthy that clustering is performed on the covariates in the training set to capture any potential dependency of the outcomes on the covariates. In situations where the relationship between the covariates and the outcome is not well-defined, the accuracy of the hard assignment approach may be compromised. Conversely, the soft assignment approach has the potential to correct for such misspecifications due to the simplified nature of the classification tree. Furthermore, this method represents an improvement over the simple averaging of cluster effects as it assigns varying importance to the established clusters. Consequently, for simulation and implementation purposes, we adopt the soft assignment approach.
\vspace{-2mm}
\section{Simulation studies}\label{simu_results}
In our simulation studies, we compare the performance of the following algorithms: Ordinary Least Squares (OLS), Linear Mixed Effects Models with Gaussian cluster effects (LMM (Gaussian)), Linear Mixed Effects Models with Generalized Laplace cluster effects (LMM (Laplace)), and our Mixed Integer Optimization approach (MIO). The Gaussian kernel has form $e^{-cx^2}$, while the Laplace kernel, with $e^{-c|x|}$, is often used as a heavy-tailed alternative to the Gaussian in the presence of outliers \citep{west1984outlier}. To impose a hierarchical or clustered structure in the simulated data, we employ a hidden confounding model as the data-generating process. We generate $X_i\sim \mathcal{N}_{P}(c^\intercal\gamma,I)$, with $c\sim \mathcal{N}_{Q}(0,I)$, for $i\in[P]$, with an intercept column, and then $Y\sim \mathcal{N}(\boldsymbol{X\beta},\sigma^2_\varepsilon)$. This approach induces a clustered structure in both the covariates and the outcomes.

We design simulations under two broad \textbf{cluster-effect types}. In Case 1, the cluster effects ($\gamma_k$s) truly originate from a Gaussian distribution, representing a classical cluster effect problem where LMM (Gaussian) is expected to perform well. In Case 2, the cluster effects are truly sparse (mostly zeroes, with a certain percentage of entries being non-zero). This scenario may not be best suited to LMM (Gaussian), yet it reflects realistic conditions. Within each type, we also vary the level of \textbf{heterogeneity}. For the sparse cases, we directly vary the sparsity (percentage of non-zero entries) of the underlying $\gamma$ vector, generating the non-zero values as ${+1,-1}$ with equal probability. For the Gaussian case, we draw $\gamma_k$'s from $\mathcal{N}(0,\sigma_\gamma^2)$ and vary the value of $\sigma_\gamma^2$ to manipulate the heterogeneity of $\gamma$. To evaluate scalability, we conduct simulations across three different levels of \textbf{data complexity} (dimensionality): ``Low'' ($k=4, P=10, Q=0$), ``Medium'' ($k=10, P=25, Q=0$), and ``High'' ($k=14, P=35, Q=0$), where $k$ represents the number of clusters, $P$ the number of non trivial covariates, and $Q$ the number of cluster-specific covariates of higher order. At each level, the number of observations within each cluster is set to $50$, ensuring a constant ratio of $p/n$ to maintain the aspect ratio'' of the problem.

In total, this design represents 66 different combinations of cluster-effect type, heterogeneity, and complexity. Although our simulations are restricted to a standard cluster-based intercepts model, the cluster-based methods can be generalized to higher-order cluster effects. We partition the data into training, validation, and testing sets, and compare the performance of these models.

\subsection{Recovery of causal effects}

\begin{figure*}[t]
\centering
    \includegraphics[width = \textwidth]{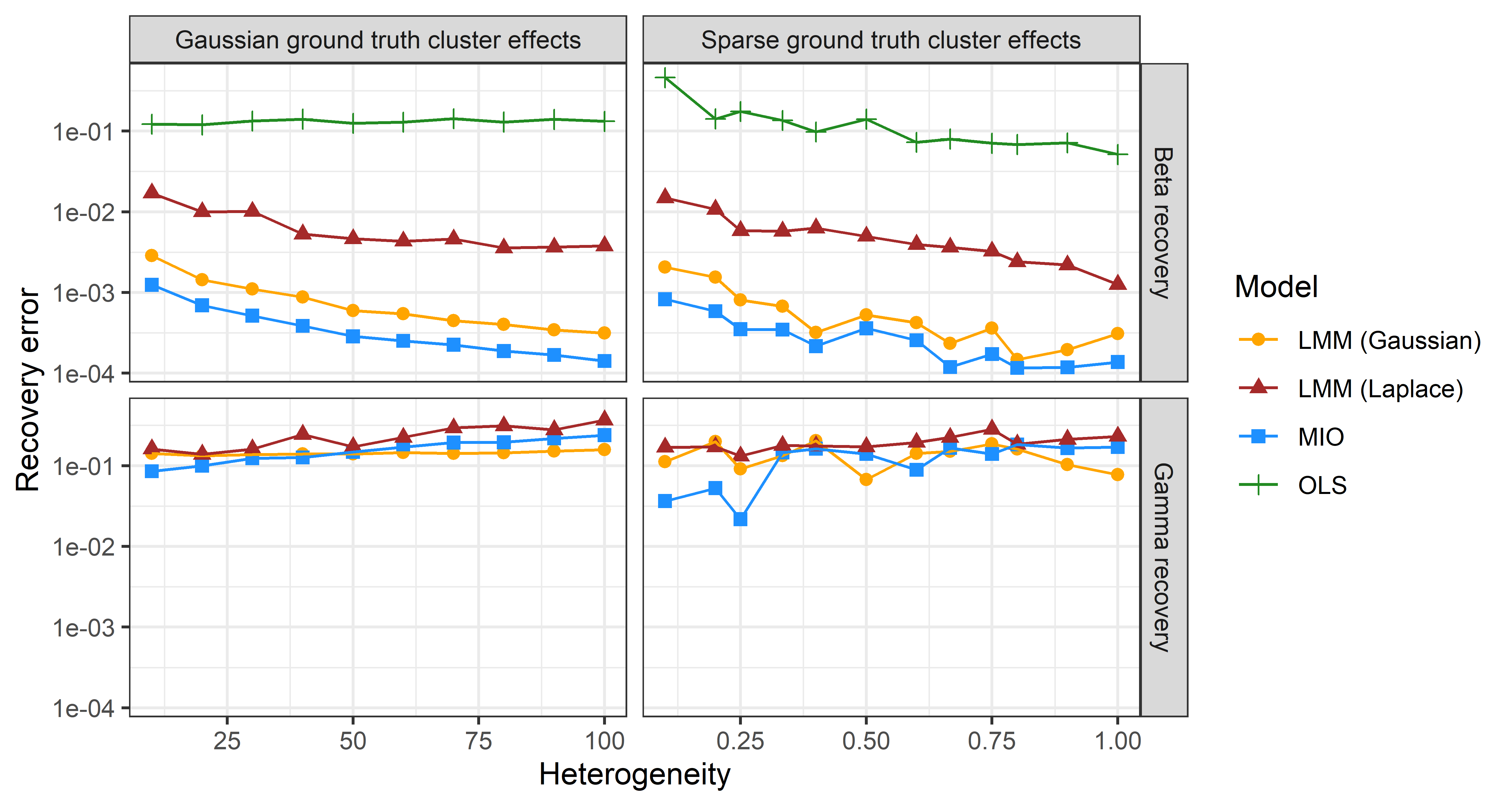}
    \caption{$\ell_2$ causal effect recovery on the log scale ($\beta$ (above) and $\gamma$ (below)) under the simulation scenarios where the cluster-effects are truly Gaussian (left) and truly sparse (right), with high dimensionality (14 clusters, 35 covariates, 50 observations per cluster)}
    \label{fig:causal_recovery_high}
\end{figure*}

An important characteristic of our method is its performance as an inferential tool. We compare the recovery of $\beta$ and $\gamma$ in a cluster-based intercept model, under all the different simulation scenarios described above. Additionally, we look at the recovery of the intra-cluster correlation (ICC), an important statistic as it codifies the inherent relatedness of observations within each cluster \citep{Bartko1966-oo}, in the Supplement.

Figure \ref{fig:causal_recovery_high} shows the results of our simulations in a ``High'' dimensional setting. The recovery here is the $\ell_2$ recovery of either $\beta$ or $\gamma$ (i.e. $\ell_2$ difference between the estimated $\beta,\gamma$ vector from each method and the ground-truth). First, we look at results under all the scenarios where the cluster-effects are truly Gaussian, as this is exactly the assumed underlying data-generation process for LMM (Gaussian). Notably, OLS and LMM (Laplace) are not performing as well as the other two algorithms. As expected, we observe LMM (Gaussian) to perform well in this setup, but we also discover that our method works at a very commensurate (and often superior) level, showing that our sparsity-constrained MIO formulation is able to flexibly adapt even when subjected to datasets that are more aligned with LMM (Gaussian).We observe that the MIO in fact has better $\beta$ recovery than LMM (Gaussian) across all possible scenarios, and found this to uniformly hold even in the ``Low'' and ``Medium'' dimensional setups, which are in the Supplement.

For the three methods that do perform cluster-effect adjustment (LMM (Gaussian), LMM (Laplace), MIO), we also looked at how well the cluster-effects ($\gamma_k$) themselves were recovered, as these are of scientific interest. We found that generally, the  methods perform similarly well, which again corroborated our belief in the MIO formulation to still perform in a setup suited for LMM (Gaussian).

Next, we investigated simulation scenarios where the cluster effects were truly sparse, since this represents a deviation from the restrictive distributional assumptions of LMM (Gaussian). Even in this case, OLS and LMM (Laplace) struggle to recover the underlying causal vector. We found notable superiority of the MIO approach compared to the other methods in recovery of the $\beta$ vector, under the ``High'' dimensional sparse setup. This demonstrates the efficacy of the MIO approach over a broad range data-generation scenarios, making it a very applicable tool. We see similar results in ``Medium'' and ``Low'' settings, given in the Supplement. Similar to the Gaussian case, we see that the methods perform similarly well in recovery of the cluster-effects ($\gamma_k$), with MIO performing much better at higher levels of sparsity, which is observed in the ``Low'' and ``Medium'' dimensionality cases as well, given in the Supplement.

In the Supplement, we have tables demonstrating the sparsity recovery of the $\gamma$ vector  and the recovery of the ICC. The MIO approach works very impressively in sparsity recovery. In the fringe cases where sparsity is very high, the MIO approach tends to take a conservative sparsity estimate. Additionally, the sparsity recovery improves with dimensionality. For the ICC, as expected, LMM (Gaussian) achieves near perfect recovery, but the MIO approach is very close as well.
\vspace{-2mm}
\subsection{Predictive performance}
\vspace{-1mm}
\begin{figure}[t]
\centering
    \includegraphics[width = 0.90\textwidth]{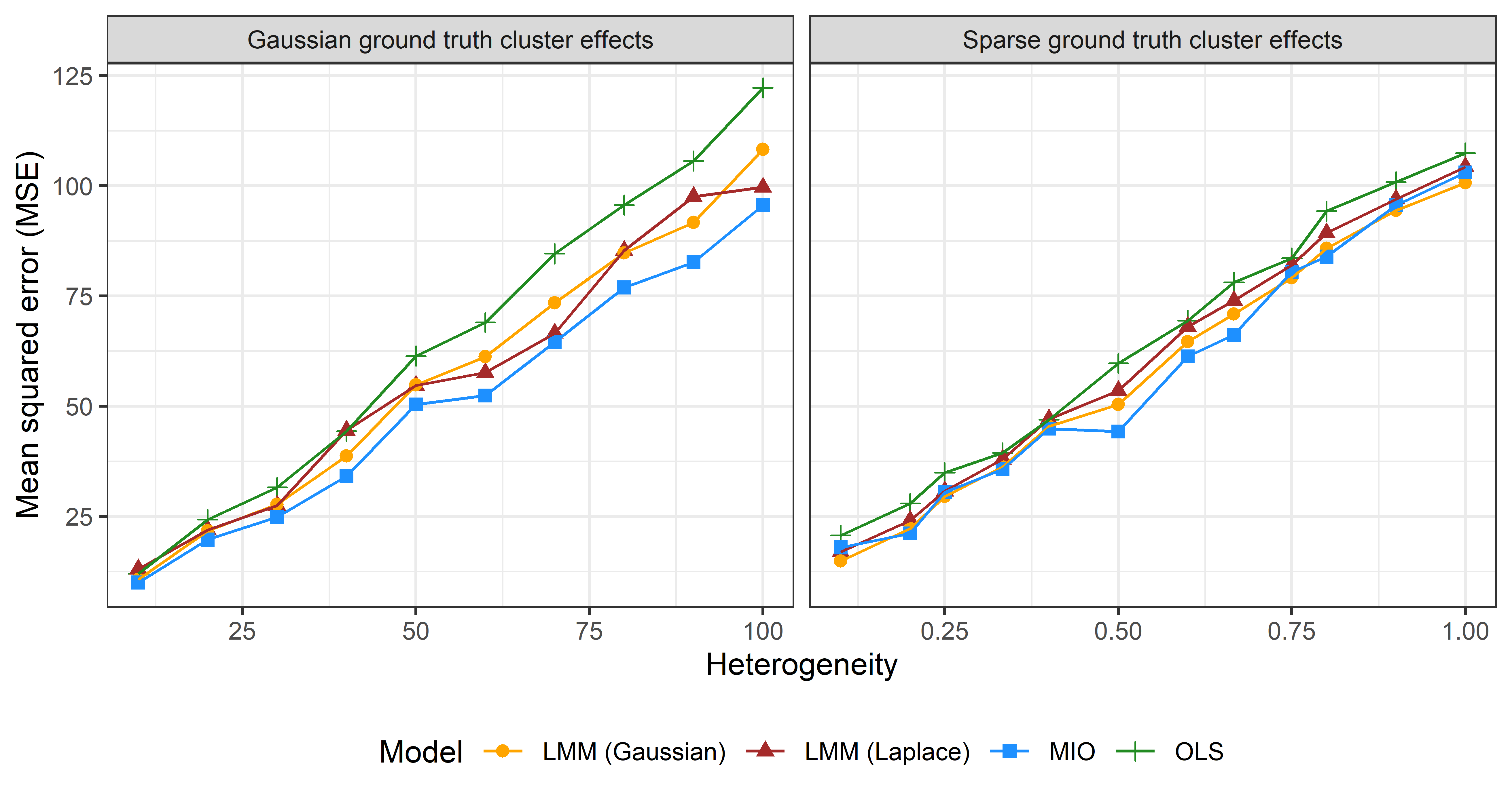}
    \caption{Predictive MSE of the four algorithms in a high dimensional setup (14 clusters, 35 covariates, 50 observations per cluster), where the cluster-effects are truly Gaussian (left) and truly sparse (right) }
    \label{fig:mse_1_high}
\end{figure}

Another important characteristic of our method is its efficacy as a predictive tool. As we elucidated in the methods section, our procedure leverages the clustered nature of the data, using classification trees, to provide insight into new data points. We compare the predictive performances of these algorithms by looking at the mean-squared error of prediction in a held-out test set, under all the previously mentioned simulation setups. Figure \ref{fig:mse_1_high} shows that the predictive MSE of the MIO method is clearly lower than that of the other methods in the Gaussian case, and has generally better performance in the sparse case, for all the scenarios under the ``High'' dimensional setup. A similar trend was observed in ``Low'' and ``Medium'' dimensions, with MIO performing uniformly better than the other algorithms, given in the Supplement. Hence, by leveraging the estimated $\gamma_k$s, we can achieve more accurate \textit{cluster-informed} predictions.
\vspace{-2mm}
\subsection{Computational cost}
\vspace{-1mm}
While conducting the simulations for parameter recovery, we additionally ran simulations to compare the computation time for the algorithms we are investigating. These simulations were run on a Macbook Pro with an Apple M1 Pro processing chip. Table \ref{Tab:comp_time} shows the average computational time in different regimes for the the different algorithms. We can see that for higher levels of specification, and in term, lower levels of restriction, the algorithm does take a longer time to converge. In particular, the generalized Laplacian LMM model takes the longest to run, in a similar order of magnitude to MIO. It is to be noted that both of these methods can actual enforce sparsity, although LMM (Laplace) requires thresholding. Thus, we can conclude that there is a price to pay in computational cost for effectively dealing with the SCS problem. 
\begin{table}[ht]
\vspace{-4mm}
\caption{Comparison of computation complexity in simulations of regression methods under different levels of dimensionality (low, medium, high) under a Sparse generation process}
\vspace{3mm}
\centering
\begin{tabular}{ |c|c|c|c|c|} 
\hline
Dimensionality &  \multicolumn{4}{c|}{Average runtime per model run (ms)}\\
\cline{2-5}
& OLS  & LMM (Gaussian)  & LMM (Laplace) &  MIO\\
\hline
\text{Low}& 2.63& 13.67& 744.55&237.22\\ 
\text{Medium}& 16.24& 18.07& 835.42& 328.47\\ 
\text{High}& 26.74& 43.07& 927.88& 539.35\\ 
\hline
\end{tabular}
\label{Tab:comp_time}
\vspace{-6mm}
\end{table}

\section{Data examples} 
\label{data_example}

\vspace{-2mm}
\subsection{Student performance}
\vspace{-1mm}
In this application, we aim to predict the academic performance of 382 Portuguese secondary school students, as documented in a study by \citet{cortez2008using}, with data sourced from the UCI Machine Learning Repository. The dataset consists of 30 predictors, including demographic, social, and school-related factors such as parental education and study time. Continuous predictors were standardized, and dummy variables were utilized for categorical predictors, resulting in a total of 20 predictors. The outcome variables are the students' test scores in Mathematics and Portuguese.

It is essential to note that in this context, a cluster effect represents the inherent aptitude of each student, considering other societal factors. Essentially, it provides a performance metric, \textbf{given} the stratum of covariates to which a student belongs. This allows the school to identify students who are over-performing or under-performing and provide them with the necessary opportunities or guidance. Therefore, our objective in this example is to recover the random intercepts using LMMs and MIO.

A simplistic approach to identifying these exceptional students might involve ranking the students' mean raw scores and identifying the students at the extremes. We will show that this regression-less method may overlook certain students since it does not account for environmental factors. To assess the concordance of the algorithms, we first compare the estimated $\boldsymbol{\beta}$ vectors across the different methods, and find them to generally agree (table in the Supplement). Hence we can proceed with the comparison of cluster effects.

\begin{figure}[h]
	\captionsetup[subfigure]{oneside,margin={0cm,0cm}}
	{\includegraphics[width = 0.471\textwidth]{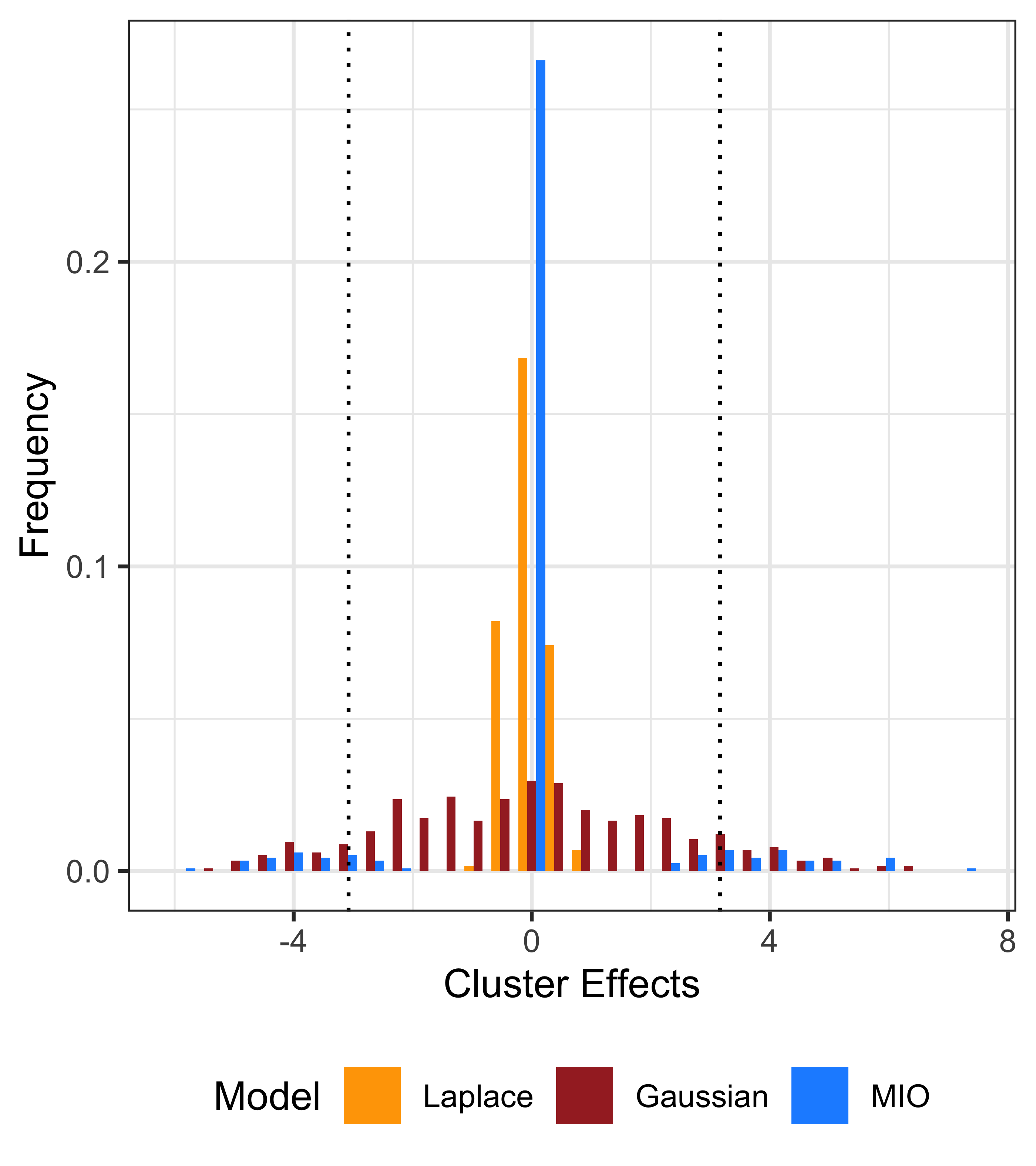}}
	\hspace{2mm}
	\captionsetup[subfigure]{oneside,margin={0cm,0cm}}
	{\includegraphics[width = 0.529\textwidth]{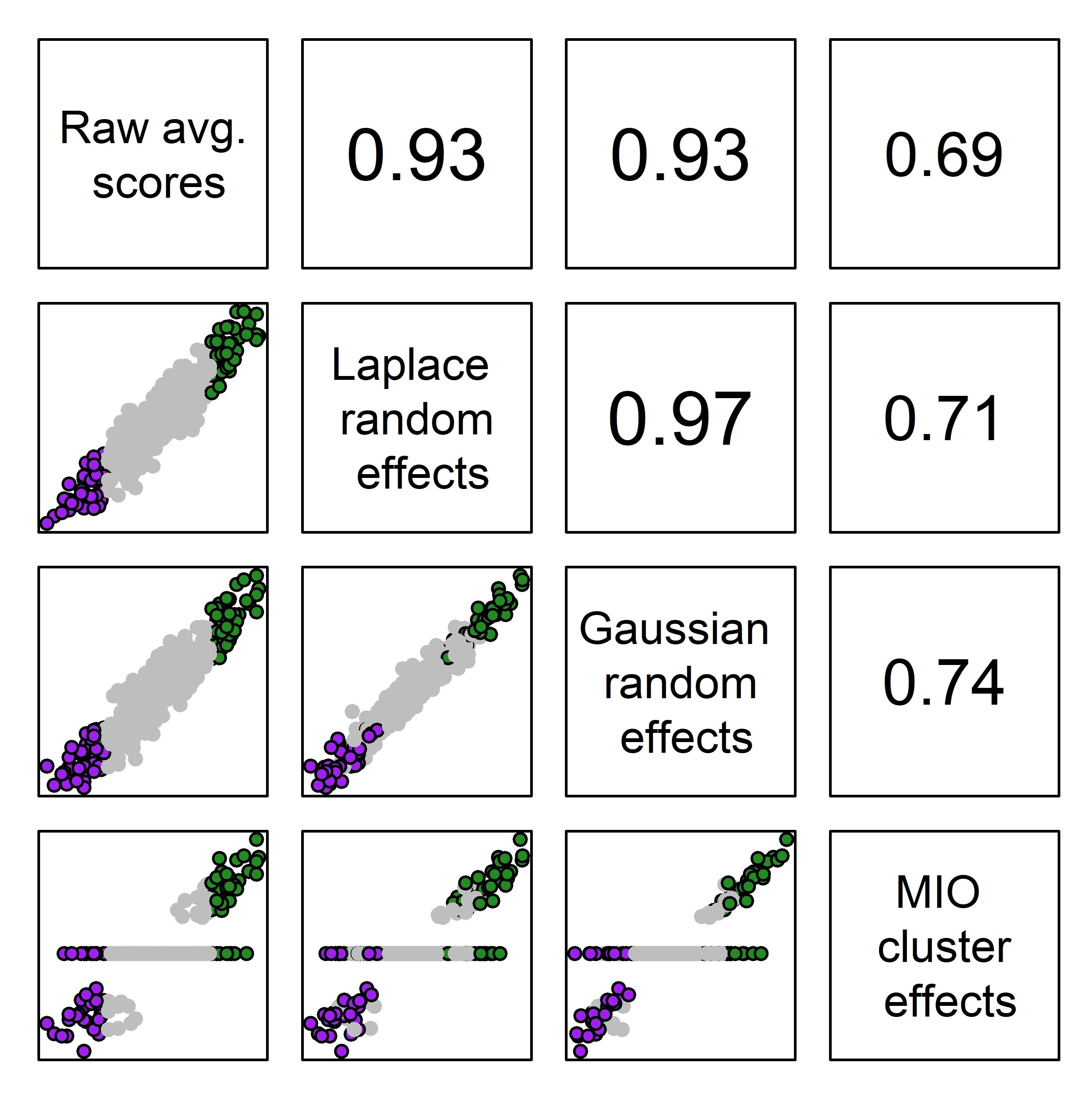}}
	\hspace{2mm}
	\caption{(Left) Predicted random effects assuming Laplace-distributed, Gaussian-distributed, and MIO and (Right) Concordance of cluster effects with raw scores across different algorithms. The points are colored based on the raw average scores (green = top 10\%, purple = bottom 10\%)} \label{fig:gamma_school} 
 \vspace{3mm} 
\end{figure}

Figure \ref{fig:gamma_school} illustrates the spread of the predicted random intercepts for each student, derived from Laplace, Gaussian, and MIO frameworks. The MIO framework reveals a distinct separation between zero and non-zero random effects, with the non-zero random effects being more definitively distanced from zero than those generated by both Laplace and Gaussian LMMs. Intriguingly, despite the Laplace distribution's propensity for variable selection, this particular instance exhibits more shrinkage than the Gaussian model. This occurrence is primarily attributable to the \texttt{nlmm} package's questionable implementation of best linear predictions for random effects, as opposed to median predictions.

A key question to address involves the concordance between students identified at the tail ends by these three methods. Figure \ref{fig:gamma_school} presents QQ-plots showcasing pairwise comparisons among the Gaussian, Laplace, and MIO methods. The QQ-plots comparing MIO with Gaussian and Laplace reveal distinct non-monotonic relationships at the tails. This suggests that MIO identifies a markedly different group of high-performing and low-performing students compared to either the Gaussian or Laplace models, both of which exhibit significant agreement with one another. The green points in all QQ-plots represent high-scoring students based solely on raw scores, that is, without adjustments for covariates. Although both Gaussian-fitted and Laplace-fitted random effects demonstrate high concordance with raw scores, MIO contrasts significantly. The MIO model assigns zero to some students with extreme scores (represented by purple or green projected onto the horizontal line), indicating that after adjustments, these students were neither exceptionally good nor bad. Conversely, some students who were originally within the middle range are identified as outliers (gray points sent to the high and low clusters), highlighting the different insights gained through the MIO model.
\vspace{-3mm}

\subsection{Protein expression}
\vspace{-1mm}
In this application, we look a protein expression dataset for mice (\citet{Dua:2019}; \citet*{Ahmed_Dhanasekaran_Block_Tong_Costa_Stasko_Gardiner_2015}), and a corresponding analysis (\citet{Higuera_Gardiner_Cios_2015}). We use the dataset to test the predictive capacity of our algorithm against the baselines we have discussed in this paper. Naïve methods such as OLS fail to capture the underlying cluster structure, and the MIO procedure is able to leverage said structure for improved prediction, while an LMM approach \textbf{fails to converge}. The full data example, with the results are in the Supplement.

\section{Discussion}

\subsection{Comparative performance}
\vspace{-1mm}
Our primary objective was to devise an algorithm that functions as a tool for both inference and prediction. To evaluate the inferential capabilities of our algorithm, we implemented simulation studies under various data generation schemes. From an inferential standpoint, our principal targets of interest were the regression coefficients $\beta$ and the random effects $\gamma$.

Our algorithm demonstrates robust performance in recovering $\beta$, even when the underlying generation process deviates from the MIO process. This suggests that the impact of dependence between cluster effects and covariates, which may impede the recovery of other inferential targets in the LMM (Gaussian), is effectively mitigated. For $\gamma$ recovery, our algorithm performs well when the random effects exhibit higher sparsity or lower variance. As expected, conditions of very low sparsity or high variance can confound the model, making simple sparsity constraints less competitive against overarching distributional assumptions.

The flexible framework of our algorithm allows for effective recovery of the underlying sparsity, with the sparsity hyperparameter selected using a validation set. This feature also contributes to a reliable inference of the intra-cluster correlation. From a predictive standpoint, our simulation studies and real-world dataset application indicate that our method consistently outperforms the other tested methods, underscoring the value of cluster-informed predictions.
\vspace{-2mm}
\subsection{Interpretability}
\vspace{-1mm}
Interpretability is a guiding principle in the development of our approach. In practice, understanding the underlying effects of different clusters and how an observation's covariates determine its cluster is crucial. Our framework, which incorporates observation-wise cluster recovery and interpretable classification trees, provides this key information.

The flexibility of our approach permits trade-offs between interpretability and predictive or inferential accuracy. If the primary objective is to assign new observations to clusters, the complexity of the classification tree can be adjusted to yield more comprehensible results. Alternatively, if the primary goal is to predict outcomes for new observations, stronger classification algorithms such as Random Forests and XGBoost could be employed at the expense of interpretability.
\vspace{-2mm}
\subsection{Regression extensibility}
\vspace{-1mm}
The versatility of the MIO approach is a notable strength, as it seamlessly integrates additional regression constraints into the model. The MIO formulation can be extended to cover many more constraints. For instance, a more elaborate MIO setup can control the sparsity and robustness of $\beta$, explicitly manage multicollinearity, and introduce regularized non-linear effects. Hence, our method bridges hierarchical models, prevalent in healthcare and economics, with the field of ``holistic'' regression \citep{Bertsimas_Li_2020}.
\vspace{-2mm}
\subsection{Future work}
\vspace{-1mm}
Our method is predicated on the assumption that cluster labels for the training set are known. A natural extension would be to handle clustering in the training set without explicit cluster assignments. Additionally, we aim to explore bootstrapping techniques to facilitate uncertainty quantification for causal effects, thus enhancing the inferential capacity of our approach. We also notice that our model can be repurposed for outlier detection. In a scenario where each cluster contains a single observation and a sparsity limit of $\tau\%$ is applied, the model would return the trimmed regression coefficient with $(100-\tau)\%$ inlier points. Existing works such as \citet*{Chin_Kee_Eriksson_Neumann_2016}, \citet{Gomez_2021}, and \citet{Jammal} have also formalized outlier detection techniques using mixed-integer optimization.
\vspace{-2mm}
\subsection{Code availability}
\vspace{-1mm}

The codebase for this project is written using Julia and available on \href{https://github.com/Madhav1812/cluster-mio}{GitHub}. The repository contains code for:
\begin{itemize}
    \itemsep0em 
    \item simulator for data-generation processes
    \item code to evaluate performance of methods on the simulated data
    \item core functions for running our MIO model on the reader's own data
\end{itemize}
 \newpage

\bibliographystyle{jneurosci}
\bibliography{cemio_bib}

\end{document}


\maketitle


\renewcommand{\thesection}{A}
\renewcommand{\theHsubsection}{A\thesection}
\section{Theoretical results}

The convergence of our algorithm can be proven using results from previous works on best subset selection. For ease of interpretation, the following results will hold for the random intercepts model described in Section 2.1. The results are generalizable and can be extended for a generic number of random effects as well. 

\begin{definition}
    If $\boldsymbol{B}$ is a matrix, we define $\boldsymbol{B}^\intercal \boldsymbol{B}$ to be the corresponding Gram matrix.
    \end{definition}

\begin{definition}
    The spectrum of a square matrix $\bs{B} \in \mathbb{R}^{n\times n}$ is the set of eigenvalues of $\{\lambda_i(\bs{B})\}_{i=1}^{n}$, where $\lambda_1 \ge \cdots \ge \lambda _n$. By definition, we always take $\lambda_1(\mathbf{B})$ to be the largest eigenvalue of $\mathbf{B}$.
\end{definition}

In order to tackle the main convergence results, we must establish some details regarding the properties of the design matrix, specifically, the spectrum of its corresponding Gram matrix.

\begin{proposition}
    Suppose $\lambda_{1}(\mathbf{X}^\tr\mathbf{X}) = \mathcal{O}(n)$, where $n = \sum_{k=1}^{K}n_k$. Then $\lambda_1(\bs{\widetilde{X}}^\intercal\bs{\widetilde{X}}) = \mathcal{O}(n)$.
\end{proposition}

\begin{proof}
    First, note that $\mathbf{A}^\tr \mathbf{A} = \text{diag}(n_1, \cdots, n_K)$, since $\mathbf{A}$ mimics an incidence matrix with individuals representing vertices and clusters representing edges in a hypergraph, and therefore $\mathbf{A}^\tr \mathbf{A}$ enumerates the size of each cluster. Next, let
    \begin{align*}
    \mathbf{N} =
        \begin{bmatrix}
         \bs{X}^\tr\bs{X} & \mathbf{0}\\
        \mathbf{0} & \bs{A}^\tr\bs{A} \end{bmatrix}\quad \text{and} \quad \mathbf{R} = \begin{bmatrix}
        \mathbf{0} & \bs{X}^\tr\bs{A}\\
        \bs{A}^\tr\bs{X} & \mathbf{0}
    \end{bmatrix}
    \end{align*}
    Then by Weyl's inequality,
    \begin{align*}
        \lambda_1(\bs{\widetilde{X}}^\tr\bs{\widetilde{X}}) \le \lambda_1(\mathbf{N}) + \lambda_1(\mathbf{R})
    \end{align*}
    Due to the block-diagonal structure of $\mathbf{N}$, it immediately follows that the spectrum of $\mathbf{N}$ is the union of the spectrums of $\mathbf{X}^\tr \mathbf{X}$ and $\mathbf{A}^\tr \mathbf{A}$, and therefore $\lambda_1(\mathbf{N}) = \max\{\mathcal{O}(n),\max\{n_1, \cdots, n_K\}\} = \mathcal{O}(n)$.
    
    Note that the eigenvalues of $\mathbf{R}$ are precisely the singular values of $\mathbf{A}^\intercal\mathbf{X}$ (with additional multiplicity); denote $\sigma_i(\mathbf{A}^\tr \mathbf{X})$ as these singular values, also in decreasing order $\sigma_1 \ge \cdots \ge \sigma_{\min\{K, P+1\}}$. We then have
    \begin{align*}
        \sigma_1(\mathbf{A}^\intercal\mathbf{X}) \le \sigma_1(\mathbf{A})\sigma_1(\mathbf{X}) = \sqrt{\lambda_1(\mathbf{A}^\tr \mathbf{A}) \lambda_1(\mathbf{X}^\tr \mathbf{X})} = \sqrt{\max\{n_1, \cdots, n_K\}\cdot\mathcal{O}(n)} = \mathcal{O}(n)
    \end{align*}
\end{proof}

In order to introduce the following theorem, we need the following setup for the optimization problem. Let us assume that the optimization problem is of the form:

$$\min_{\boldsymbol{\beta}}g(\bs{\beta})\text{ subject to }||\bs{\beta}||_0\leq k$$
where $g$ is a convex function, and has a $\ell$-Lipschitz gradient, that is 
$$||\nabla g(\bs{\beta})-\nabla g(\bs{\widetilde{\beta}})||\leq \ell||\bs{\beta}-\bs{\widetilde{\beta}}||$$

In the case of the least squares setup we have, the explicit form of $\ell$ is given by $\lambda_1(\bs{X}^\tr\bs{X})$, where $X$ is the design matrix. Thus, from the previous proposition, the Lipschitz constant grows at the same rate as a normal regression of $Y$ on $\bs{X}$.

Now, the approach to solving this problem for a general convex function $g$ is to use an upper bound of simpler form for the function, and minimize over this upper bound. This is the essence of the outer approximation algorithm. The construction of these upper bounds requires a variable $L>\ell$ which represents the coarseness of the approximation, and in term, the ``step size'' of the descent algorithm (larger $L$, finer approximation). 

Thus, under the algorithm given in \cite{Bertsimas_King_Mazumder_2015}, the following theorem follows.

\begin{theorem}[Theorem 3.1, \cite{Bertsimas_King_Mazumder_2015}]
    Let $L > \ell$ and $\bs{\beta}^*$ be the optimum of the optimization function. After $M$ iterations, we have
    \begin{align*}
        \min_{m = 1, \cdots, M} \|\bs{\beta}_{m+1} - \bs{\beta}_m\|^2_2 \le \frac{2(g(\bs{\beta}) - g(\bs{\beta}^*))}{M(L-\ell)}
    \end{align*}
\end{theorem}

The convergence rate of our algorithm is contingent on the complexity of the function $g$ and the magnitude of $\ell$. According to our proposition, we have adjusted the growth rate of $\ell$ to align with the rate provided in the associated paper, with $g$ representing the straightforward least squares error function.

The underlying principle of Cauchy convergence implies a trajectory towards the optimum solution, given that we consider a restricted complete support for our vector. This results in a framework of convergence for our algorithm towards a global optimum.

Furthermore, we wish to highlight the efficiency of the algorithm in retrieving non-zero coefficients, also referred to as the "active set". The algorithm, by design, initializes the first $p$ coordinates as non-sparse, along with $\lambda$ proportion of the remaining coordinates. This strategy enables the algorithm to utilize outer approximation cuts to accurately identify the real active set, followed by a phase of refining the effect sizes. Empirical observations corroborated this phenomenon, providing a rationale for limiting the algorithm's iterations as a means to curtail computation time.


\newpage 
\renewcommand{\thesection}{B}
\renewcommand{\theHsubsection}{B\thesection}
\section{Simulation results}

We provide further simulation results that were not included in the main body of the paper. All results are presented in tabular form as well. The results corroborate the conclusions we drew in the main body of the paper. In particular, compared to the other approaches, the MIO approach is generally more effective in the recovery of the causal parameters of interest ($\boldsymbol{\beta}$ and $\boldsymbol{\gamma}$), and can subsequently leverage the parameters for improved prediction (i.e. better MSE on the test set), and inference (i.e. correct recovery of sparsity/ICC).

\renewcommand\thetable{\thesection\arabic{table}}  
\setcounter{table}{0}
\renewcommand\thefigure{\thesection\arabic{figure}}  
\setcounter{figure}{0}


\subsection{Cluster effects are truly sparse}
\subsubsection{$\beta$ recovery}

\begin{table}[h!]
\caption{Beta recovery under the simulation scenarios where the cluster-effects are truly sparse, with low dimensionality (4 clusters, 10 covariates, 50 observations per cluster)}

\centering
\begin{tabular}{|c|c|c|c|c|} 
\hline
True sparsity in $\gamma_k$ & $||\beta_{\text{true}} - \widehat\beta_{OLS}||_2^2$ & $||\beta_{\text{true}} - \widehat\beta_{LMM(G)}||_2^2$  & $||\beta_{\text{true}} - \widehat\beta_{LMM(L)}||_2^2$ &$||\beta_{\text{true}} - \widehat\beta_{MIO}||_2^2$ \\

\hline
        90\% & 0.23564 & 0.00122 & 0.01664&\textbf{0.00054}\\ 
        80\% & 0.23334 & 0.00116 & 0.01381&\textbf{0.00055}\\ 
        75\% & 0.23536 & 0.00121 & 0.01248&\textbf{0.00052}\\ 
        66\% & 0.10705 & 0.00056 & 0.00981&\textbf{0.00024}\\ 
        60\% & 0.10894 & 0.00060 & 0.00781&\textbf{0.00028}\\ 
        50\% & 0.10477 & 0.00061 & 0.00546&\textbf{0.00029}\\ 
        40\% & 0.07245 & 0.00038 & 0.00310&\textbf{0.00019}\\ 
        33\% & 0.07114 & 0.00036 & 0.00167&\textbf{0.00018}\\ 
        25\% & 0.07129 & 0.00035 & 0.00098&\textbf{0.00018}\\ 
        20\% & 0.04846 & 0.00025 & 0.00083&\textbf{0.00013}\\ 
        10\% & 0.05316 & 0.00030 & 0.00071&\textbf{0.00014} \\ 
        0\% & 0.05406 & 0.00024 & 0.00052&\textbf{0.00013}\\
\hline
\end{tabular}
\label{tab:beta_rec_1_low}

\end{table}

\begin{table}[h!]
\caption{Beta recovery under the simulation scenarios where the cluster-effects are truly sparse, with medium dimensionality (10 clusters, 25 covariates, 50 observations per cluster)}

\centering
\begin{tabular}{|c|c|c|c|c|} 
\hline
True sparsity in $\gamma_k$ & $||\beta_{\text{true}} - \widehat\beta_{OLS}||_2^2$ & $||\beta_{\text{true}} - \widehat\beta_{LMM(G)}||_2^2$   & $||\beta_{\text{true}} - \widehat\beta_{LMM(L)}||_2^2$ &$||\beta_{\text{true}} - \widehat\beta_{MIO}||_2^2$ \\

\hline
90\% & 0.29369 & 0.00402 & 0.01492&\textbf{0.00143}\\ 
        80\% & 0.28198 & 0.00159 & 0.00877&\textbf{0.00069} \\ 
        75\% & 0.18567 & 0.00099 & 0.00462&\textbf{0.00045}\\ 
        66\% & 0.13559 & 0.00073 & 0.00436&\textbf{0.00034} \\ 
        60\% & 0.14095 & 0.00071 & 0.00419&\textbf{0.00035} \\ 
        50\% & 0.11050 & 0.00056 & 0.00477&\textbf{0.00026} \\ 
        40\% & 0.09078 & 0.00045 & 0.00467&\textbf{0.00022}\\ 
        33\% & 0.07682 & 0.00038 & 0.00342&\textbf{0.00018}\\ 
        25\% & 0.06779 & 0.00034 & 0.00326&\textbf{0.00017}\\ 
        20\% & 0.06986 & 0.00034 & 0.00338&\textbf{0.00016}\\ 
        10\% & 0.05724 & 0.00029 & 0.00317&\textbf{0.00014}\\ 
        0\% & 0.05325 & 0.00027 & 0.00296&\textbf{0.00014}\\ 
\hline
\end{tabular}
\label{Tab:beta_rec_1_med}
\end{table}

\begin{table}[t]
\caption{Beta recovery under the simulation scenarios where the cluster-effects are truly sparse, with high dimensionality (14 clusters, 35 covariates, 50 observations per cluster)}

\centering
\begin{tabular}{|c|c|c|c|c|} 
\hline
True sparsity in $\gamma_k$ & $||\beta_{\text{true}} - \widehat\beta_{OLS}||_2^2$ & $||\beta_{\text{true}} - \widehat\beta_{LMM(G)}||_2^2$   & $||\beta_{\text{true}} - \widehat\beta_{LMM(L)}||_2^2$ &$||\beta_{\text{true}} - \widehat\beta_{MIO}||_2^2$ \\

\hline
90\% & 0.45927 & 0.00205 & 0.01489&\textbf{0.00082}\\ 
        80\% & 0.14163 & 0.00154 & 0.01074&\textbf{0.00059} \\ 
        75\% & 0.17521 & 0.00080 & 0.00583&\textbf{0.00035}\\ 
        66\% & 0.13538 & 0.00067 & 0.00576&\textbf{0.00034} \\ 
        60\% & 0.09782 & 0.00032 & 0.00629&\textbf{0.00022} \\ 
        50\% & 0.14049 & 0.00053 & 0.00497&\textbf{0.00036} \\ 
        40\% & 0.07279 & 0.00042 & 0.00394&\textbf{0.00025}\\ 
        33\% & 0.07975 & 0.00023 & 0.00364&\textbf{0.00012}\\ 
        25\% & 0.07071 & 0.00036 & 0.00324&\textbf{0.00017}\\ 
        20\% & 0.06822 & 0.00015 & 0.00240&\textbf{0.00012}\\ 
        10\% & 0.07138 & 0.00019 & 0.00219&\textbf{0.00012}\\ 
        0\% & 0.05130 & 0.00031 & 0.00126&\textbf{0.00014}\\ 
\hline
\end{tabular}
\label{Tab:beta_rec_1_high}
\end{table}

\clearpage
\subsubsection{$\gamma$ recovery}

\begin{table}[h]
\caption{Gamma recovery for cluster-based methods under the simulation scenarios where the cluster-effects are truly sparse, with low dimensionality (4 clusters, 10 covariates, 50 observations per cluster)}

\centering
\begin{tabular}{ |c|c|c|c| } 
\hline
True sparsity in $\gamma_k$ &$||\gamma_{\text{true}} - \widehat\gamma_{LMM(G)}||_2^2$ &$||\gamma_{\text{true}} - \widehat\gamma_{LMM(L)}||_2^2$ &$||\gamma_{\text{true}} - \widehat\gamma_{MIO}||_2^2$ \\
\hline
90\% & 0.03988 & 0.05686&\textbf{0.02347} \\ 
        80\% & 0.04573 & 0.07390&\textbf{0.02194} \\ 
        75\% & 0.04849 & 0.06127&\textbf{0.02554} \\ 
        66\% & 0.04519 & 0.11464&\textbf{0.03990} \\ 
        60\% & 0.04687 & 0.09935&\textbf{0.03587} \\ 
        50\% & 0.04173 & 0.11481&\textbf{0.04027} \\ 
        40\% & 0.05245 & 0.19631&\textbf{0.04295} \\ 
        33\% & 0.04680 & 0.08649&\textbf{0.04345} \\ 
        25\% & 0.05033 & 0.09610&\textbf{0.04175} \\ 
        20\% & \textbf{0.04401} & 0.09123&0.05061 \\ 
        10\% & \textbf{0.03825} & 0.06293&0.05017 \\ 
        0\% & \textbf{0.04061} & 0.11586&0.05236 \\ 
\hline
\end{tabular}
\label{Tab:gamma_rec_low}
\end{table}

\begin{table}[h!]
\caption{Gamma recovery for cluster-based methods under the simulation scenarios where the cluster-effects are truly sparse, with medium dimensionality (10 clusters, 25 covariates, 50 observations per cluster)}

\centering
\begin{tabular}{ |c|c|c|c| } 
\hline
True sparsity in $\gamma_k$ &$||\gamma_{\text{true}} - \widehat\gamma_{LMM(G)}||_2^2$ &$||\gamma_{\text{true}} - \widehat\gamma_{LMM(L)}||_2^2$ &$||\gamma_{\text{true}} - \widehat\gamma_{MIO}||_2^2$ \\
\hline
        90\% & 0.11046 & 0.08674&\textbf{0.03191} \\ 
        80\% & 0.09521 & 0.07189&\textbf{0.05322} \\ 
        75\% & 0.10700 & 0.14838&\textbf{0.06776} \\ 
        66\% & 0.10952 & 0.15673&\textbf{0.07681} \\ 
        60\% & 0.10673 & 0.14789&\textbf{0.07519} \\ 
        50\% & 0.10657 & 0.16863&\textbf{0.08277} \\ 
        40\% & 0.11001 & 0.18258&\textbf{0.10667} \\ 
        33\% & 0.10921 & 0.16142&\textbf{0.10074} \\ 
        25\% & 0.10388 & 0.17106&\textbf{0.09744} \\ 
        20\% & \textbf{0.10748} & 0.15626&0.12749 \\ 
        10\% & \textbf{0.10500} & 0.17360&0.14658 \\ 
        0\% & \textbf{0.12681} & 0.16553&0.14561 \\ 
\hline
\end{tabular}
\label{Tab:gamma_rec_med}
\end{table}

\begin{table}[h!]
\caption{Gamma recovery for cluster-based methods under the simulation scenarios where the cluster-effects are truly sparse, with high dimensionality (14 clusters, 35 covariates, 50 observations per cluster)}

\centering
\begin{tabular}{ |c|c|c|c| } 
\hline
True sparsity in $\gamma_k$ &$||\gamma_{\text{true}} - \widehat\gamma_{LMM(G)}||_2^2$ &$||\gamma_{\text{true}} - \widehat\gamma_{LMM(L)}||_2^2$ &$||\gamma_{\text{true}} - \widehat\gamma_{MIO}||_2^2$ \\
\hline
90\% & 0.11147 & 0.16897&\textbf{0.03638} \\ 
        80\% & 0.19905 & 0.17153&\textbf{0.05270} \\ 
        75\% & 0.09119 & 0.13122&\textbf{0.02171} \\ 
        66\% & \textbf{0.13254} & 0.17776&0.14525 \\ 
        60\% & 0.20280 & 0.17549&\textbf{0.16220} \\ 
        50\% & \textbf{0.06703} & 0.17158&0.13927 \\ 
        40\% & 0.14134 & 0.19434 &\textbf{0.00887} \\ 
        33\% & \textbf{0.15031} & 0.22395&0.16627 \\ 
        25\% & 0.18617 & 0.28267&\textbf{0.13943} \\ 
        20\% & \textbf{0.15981} & 0.18452&0.18249 \\ 
        10\% & \textbf{0.10285} & 0.21238&0.16508 \\ 
        0\% & \textbf{0.07705} & 0.23086&0.16996 \\ 
\hline
\end{tabular}
\label{Tab:gamma_rec_high}
\end{table}

\clearpage
\subsubsection{Sparsity recovery}

It should be noted, that the competing methods (Laplace LMM and Gaussian LMM) cannot recovery sparse cluster effects $\gamma_k$, i.e. inferred sparsity of $\gamma_k$ is always $0\%$, regardless of what the true sparsity is. In comparison, we see efficient recovery of the sparsity my the MIO approach, and this recovery even improves with the dimensionality of the regresion problem.

\begin{table}[h!]
\caption{Sparsity recovery of Gamma of the MIO approach under the different dimensional setups}

\centering
\begin{tabular}{ |c|c|c|c| } 
\hline
True sparsity in $\gamma_k$ & Low & Medium & High \\
\hline
90\% & 79\% &82\% &86\% \\ 
80\% & 70\% &71\%  & 79\%\\ 
75\% & 60\% &65\%  & 72\%\\ 
66\% & 57\%&59\%  &63\%\\ 
60\% & 49\%&53\%  & 58\%\\ 
50\% & 41\% &43\%  & 45\%\\ 
40\% & 38\%&33\%  & 39\%\\ 
33\% & 25\%&25\%  &31\%\\ 
25\% & 15\%&15\%  &20\%\\ 
20\% & 11\%&14\% &15\%\\ 
10\% & 0\%&6\%  &8\%\\ 
0\% &0\% & 0\%  &0\%\\ 
\hline
\end{tabular}
\label{Tab:gamma_spar}
\end{table}

\subsection{Cluster effects are truly Gaussian}
\subsubsection{$\beta$ recovery}

\begin{table}[h!]
\caption{Beta recovery under the simulation scenarios where the cluster-effects are truly Gaussian, with low dimensionality (4 clusters, 10 covariates, 50 observations per cluster)}
\label{tab:beta_rec_1_low_g}
\centering
\begin{tabular}{|c|c|c|c|c|} 
\hline
Variance of $\gamma_k$ & $||\beta_{\text{true}} - \widehat\beta_{OLS}||_2^2$ & $||\beta_{\text{true}} - \widehat\beta_{LMM(G)}||_2^2$ & $||\beta_{\text{true}} - \widehat\beta_{LMM(L)}||_2^2$ &$||\beta_{\text{true}} - \widehat\beta_{MIO}||_2^2$  \\

\hline
        10.0 & 0.09206 & 0.00378 & 0.04470&\textbf{0.00175}\\ 
        20.0 & 0.10171 & 0.00199 & 0.03591&\textbf{0.00089}\\ 
        30.0 & 0.10263 & 0.00132 & 0.02300&\textbf{0.00072}\\ 
        40.0 & 0.10711 & 0.00134 & 0.01236&\textbf{0.00061}\\ 
        50.0 & 0.09984 & 0.00094 & 0.01030&\textbf{0.00047}\\ 
        60.0 & 0.10761 & 0.00111 & 0.00796&\textbf{0.00052}\\ 
        70.0 & 0.11339 & 0.00060 & 0.00682&\textbf{0.00027}\\ 
        80.0 & 0.10867 & 0.00069 & 0.00378&\textbf{0.00031}\\ 
        90.0 & 0.10310 & 0.00049 & 0.00154&\textbf{0.00026}\\ 
        100.0 & 0.10034 & 0.00045 & 0.00078&\textbf{0.00020}\\ 
\hline
\end{tabular}
\end{table}

\begin{table}[ht]
\caption{Beta recovery under the simulation scenarios where the cluster-effects are truly Gaussian, with medium dimensionality (10 clusters, 25 covariates, 50 observations per cluster)}

\centering
\begin{tabular}{|c|c|c|c|c|} 
\hline
Variance of $\gamma_k$ & $||\beta_{\text{true}} - \widehat\beta_{OLS}||_2^2$ & $||\beta_{\text{true}} - \widehat\beta_{LMM(G)}||_2^2$ & $||\beta_{\text{true}} - \widehat\beta_{LMM(L)}||_2^2$ &$||\beta_{\text{true}} - \widehat\beta_{MIO}||_2^2$  \\

\hline
10.0 & 0.11150 & 0.00289 & 0.00947&\textbf{0.00142} \\ 
        20.0 & 0.12713 & 0.00161 & 0.00656&\textbf{0.00074}\\ 
        30.0 & 0.11915 & 0.00116 & 0.00569&\textbf{0.00056}\\ 
        40.0 & 0.13113 & 0.00094 & 0.00468&\textbf{0.00043}\\ 
        50.0 & 0.12789 & 0.00068 & 0.00477&\textbf{0.00034}\\ 
        60.0 & 0.11203 & 0.00061 & 0.00463&\textbf{0.00029}\\ 
        70.0 & 0.12111 & 0.00045 & 0.00475&\textbf{0.00020}\\ 
        80.0 & 0.13204 & 0.00042 & 0.00409&\textbf{0.00020}\\ 
        90.0 & 0.12829 & 0.00042 & 0.00376&\textbf{0.00020}\\ 
        100.0 & 0.13437 & 0.00035 & 0.00429&\textbf{0.00016 }\\ 
\hline
\end{tabular}
\label{Tab:beta_rec_1_med_g}
\end{table}

\begin{table}[ht]
\caption{Beta recovery under the simulation scenarios where the cluster-effects are truly Gaussian, with high dimensionality (14 clusters, 35 covariates, 50 observations per cluster)}

\centering
\begin{tabular}{|c|c|c|c|c|} 
\hline
Variance of $\gamma_k$ & $||\beta_{\text{true}} - \widehat\beta_{OLS}||_2^2$ & $||\beta_{\text{true}} - \widehat\beta_{LMM(G)}||_2^2$ & $||\beta_{\text{true}} - \widehat\beta_{LMM(L)}||_2^2$ &$||\beta_{\text{true}} - \widehat\beta_{MIO}||_2^2$  \\

\hline
10.0 & 0.12196 & 0.00285 & 0.01704&\textbf{0.00124} \\ 
        20.0 & 0.11948 & 0.00143 & 0.01000&\textbf{0.00069}\\ 
        30.0 & 0.13324 & 0.00110 & 0.01011&\textbf{0.00052}\\ 
        40.0 & 0.14063 & 0.00087 & 0.00529&\textbf{0.00039}\\ 
        50.0 & 0.12514 & 0.00060 & 0.00463&\textbf{0.00029}\\ 
        60.0 & 0.12936 & 0.00054 & 0.00432&\textbf{0.00025}\\ 
        70.0 & 0.14265 & 0.00045 & 0.00459&\textbf{0.00023}\\ 
        80.0 & 0.12956 & 0.00040 & 0.00356&\textbf{0.00019}\\ 
        90.0 & 0.13977 & 0.00034 & 0.00364&\textbf{0.00017}\\ 
        100.0 & 0.13182 & 0.00031 & 0.00378&\textbf{0.00014 }\\ 
\hline
\end{tabular}
\label{Tab:beta_rec_1_high_g}
\end{table}

\clearpage
\subsubsection{$\gamma$ recovery}

\begin{table}[h!]
\caption{Gamma recovery for cluster-based methods under the simulation scenarios where the cluster-effects are truly Gaussian, with low dimensionality (4 clusters, 10 covariates, 50 observations per cluster)}

\centering
\begin{tabular}{ |c|c|c|c| } 
\hline
Variance of $\gamma_k$ &$||\gamma_{\text{true}} - \widehat\gamma_{LMM(G)}||_2^2$ &$||\gamma_{\text{true}} - \widehat\gamma_{LMM(L)}||_2^2$ &$||\gamma_{\text{true}} - \widehat\gamma_{MIO}||_2^2$ \\
\hline
10.0 & 0.04014 & 0.07613&\textbf{0.02846} \\ 
        20.0 & 0.04251 & 0.13257&\textbf{0.02961} \\ 
        30.0 & 0.04119 & 0.10150&\textbf{0.03812} \\ 
        40.0 & 0.04579 & 0.14589&\textbf{0.04118} \\ 
        50.0 & \textbf{0.03857} & 0.17664&0.04495 \\ 
        60.0 & \textbf{0.04116 }& 0.08313&0.04400 \\ 
        70.0 & \textbf{0.04085} & 0.08387&0.05030 \\ 
        80.0 & \textbf{0.03746} & 0.17306&0.05121 \\ 
        90.0 & \textbf{0.04510} & 0.14564&0.06128 \\ 
        100.0 & \textbf{0.04305} & 0.14890&0.06324 \\ 
\hline
\end{tabular}
\label{Tab:gamma_rec_low_g}
\end{table}

\begin{table}[ht]
\caption{Gamma recovery for cluster-based methods under the simulation scenarios where the cluster-effects are truly Gaussian, with medium dimensionality (10 clusters, 25 covariates, 50 observations per cluster)}
\centering
\begin{tabular}{ |c|c|c|c| } 
\hline
Variance of $\gamma_k$ &$||\gamma_{\text{true}} - \widehat\gamma_{LMM(G)}||_2^2$ &$||\gamma_{\text{true}} - \widehat\gamma_{LMM(L)}||_2^2$ &$||\gamma_{\text{true}} - \widehat\gamma_{MIO}||_2^2$ \\
\hline
10.0 & 0.10519 & 0.17888&\textbf{0.07696} \\ 
        20.0 & 0.09958 & 0.15849&\textbf{0.09785} \\ 
        30.0 & 0.10009 & 0.13542&\textbf{0.08733} \\ 
        40.0 & 0.10931 & 0.13914&\textbf{0.10110} \\ 
        50.0 & \textbf{0.10460} & 0.17986&0.11122 \\ 
        60.0 & \textbf{0.09965} & 0.16779&0.11120 \\ 
        70.0 &\textbf{ 0.10389} & 0.15837&0.13616 \\ 
        80.0 &\textbf{ 0.10631} & 0.17409&0.14843 \\ 
        90.0 & \textbf{0.10476} & 0.18811&0.15833 \\ 
        100.0 & \textbf{0.10682} & 0.18989&0.16284 \\ 
\hline
\end{tabular}
\label{Tab:gamma_rec_med_g}
\end{table}

\begin{table}[ht]
\caption{Gamma recovery for cluster-based methods under the simulation scenarios where the cluster-effects are truly Gaussian, with high dimensionality (14 clusters, 35 covariates, 50 observations per cluster)}
\centering
\begin{tabular}{ |c|c|c|c| } 
\hline
Variance of $\gamma_k$ &$||\gamma_{\text{true}} - \widehat\gamma_{LMM(G)}||_2^2$ &$||\gamma_{\text{true}} - \widehat\gamma_{LMM(L)}||_2^2$ &$||\gamma_{\text{true}} - \widehat\gamma_{MIO}||_2^2$ \\
\hline
10.0 & 0.14146 & 0.16100&\textbf{0.08520} \\ 
        20.0 & 0.13318 & 0.13843&\textbf{0.09959} \\ 
        30.0 & 0.13711 & 0.16041&\textbf{0.12369} \\ 
        40.0 & 0.14007 & 0.24408&\textbf{0.12649} \\ 
        50.0 & \textbf{0.13960} & 0.17184&0.14740 \\ 
        60.0 & \textbf{0.14571} & 0.22432&0.16954 \\ 
        70.0 &\textbf{ 0.14166} & 0.29360&0.19313 \\ 
        80.0 &\textbf{ 0.14382} & 0.31070&0.19481 \\ 
        90.0 & \textbf{0.15180} & 0.27851&0.21748 \\ 
        100.0 & \textbf{0.15772} & 0.36796&0.23885 \\ 
\hline
\end{tabular}
\label{Tab:gamma_rec_high_g}
\end{table}

\clearpage
\subsubsection{Intra-cluster correlation (ICC) recovery}

We analyze the Intraclass Correlation Coefficient (ICC) recovery of our algorithm in comparison to the baseline methodologies. ICC serves as a crucial parameter given its capacity to capture intra-cluster observation behavior, which has broad applications in various clinical settings, such as assessing hospital homogeneity.

Our simulations illustrate that the Linear Mixed Model (LMM) with Gaussian cluster effects accurately recovers the true ICC, which is anticipated given its alignment with the underlying data generation process. Intriguingly, our Mixed Integer Optimization (MIO) model also demonstrates commendable ICC recovery, especially when contrasted with the LMM fitted with Laplace cluster effects. The MIO exhibits enhanced recovery for higher ICC levels, signifying stronger within-group interactions. This marked improvement in the MIO's performance underscores its effectiveness in handling more complex interaction dynamics.

\begin{table}[h!]
\caption{Comparison of ICC recovery in Medium dimensionality (10 clusters, 25 covariates, 50 observations per cluster) between LMM (Gaussian) and MIO}
\centering
\begin{tabular}{ |c|c|c|c| } 
\hline
True ICC &LMM (Gaussian) ICC& LMM (Laplace) ICC&MIO ICC\\
\hline
10\% & 8\%& 1\%&4\%  \\ 
20\% & 18\%& 8\%&13\%\\ 
30\% & 29\%& 20\%&25\%\\ 
40\% & 39\%& 31\%&39\%\\ 
50\% & 49\%& 41\%&48\%\\ 
60\% & 59\%& 50\%&59\%\\ 
70\% & 70\%& 59\%&69\% \\ 
80\% & 80\%& 68\%&79\%\\ 
90\% & 90\%& 89\%&90\%\\ 
\hline
\end{tabular}
\label{Tab:icc}
\end{table}

\subsection{Further simulation plots}

\begin{figure*}[h]
\centering
    \includegraphics[width = \textwidth]{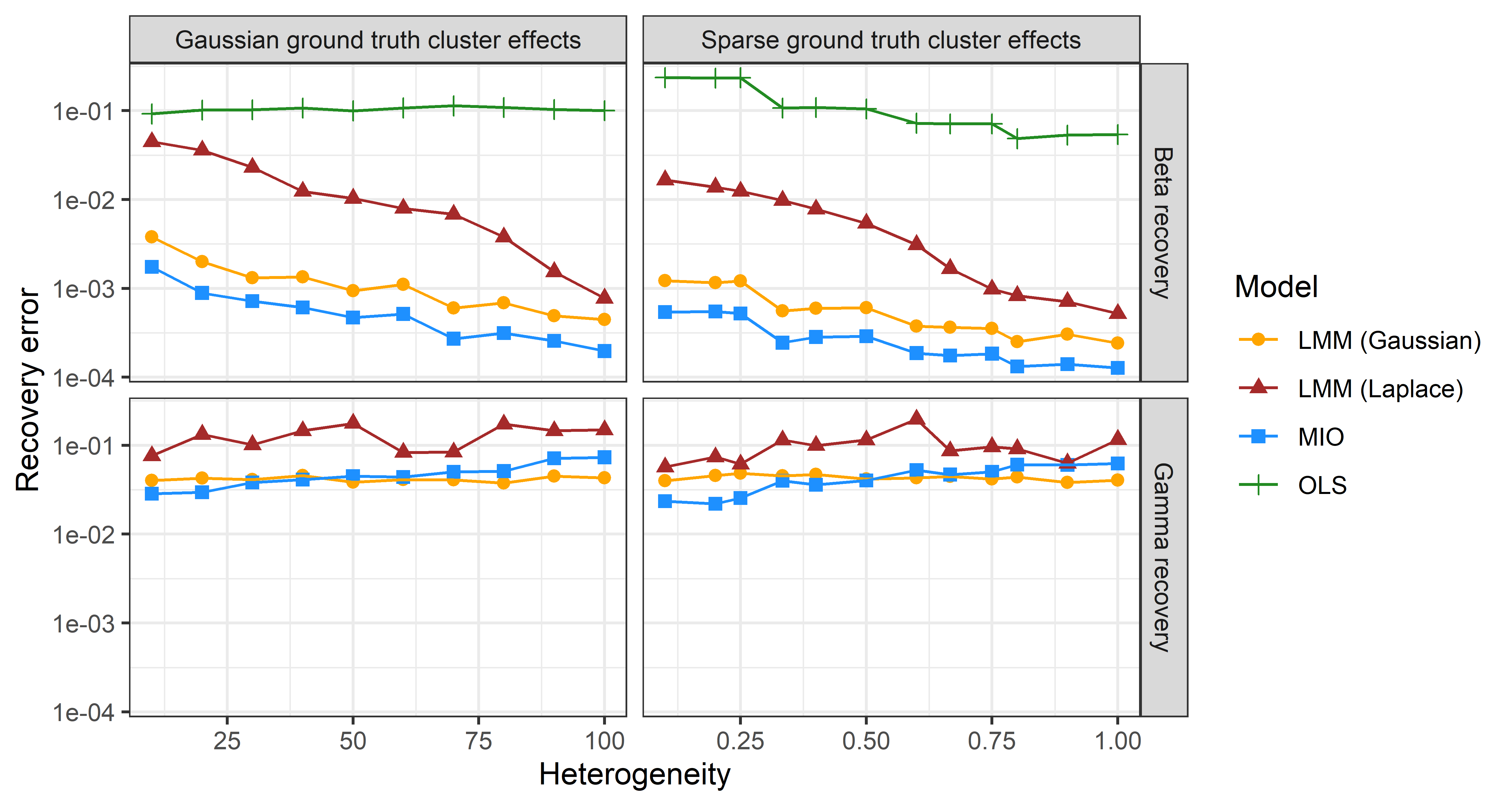}
    \caption{$\ell_2$ causal effect recovery on the log scale ($\beta$ (above) and $\gamma$ (below)) under the simulation scenarios where the cluster-effects are truly Gaussian (left) and truly sparse (right), with low dimensional setup (4 clusters, 10 covariates, 50 observations per cluster), where the cluster-effects are truly Gaussian (left) and truly sparse (right)}
    \label{fig:causal_recovery_low}
\end{figure*}

\begin{figure*}[h]
\centering
    \includegraphics[width = \textwidth]{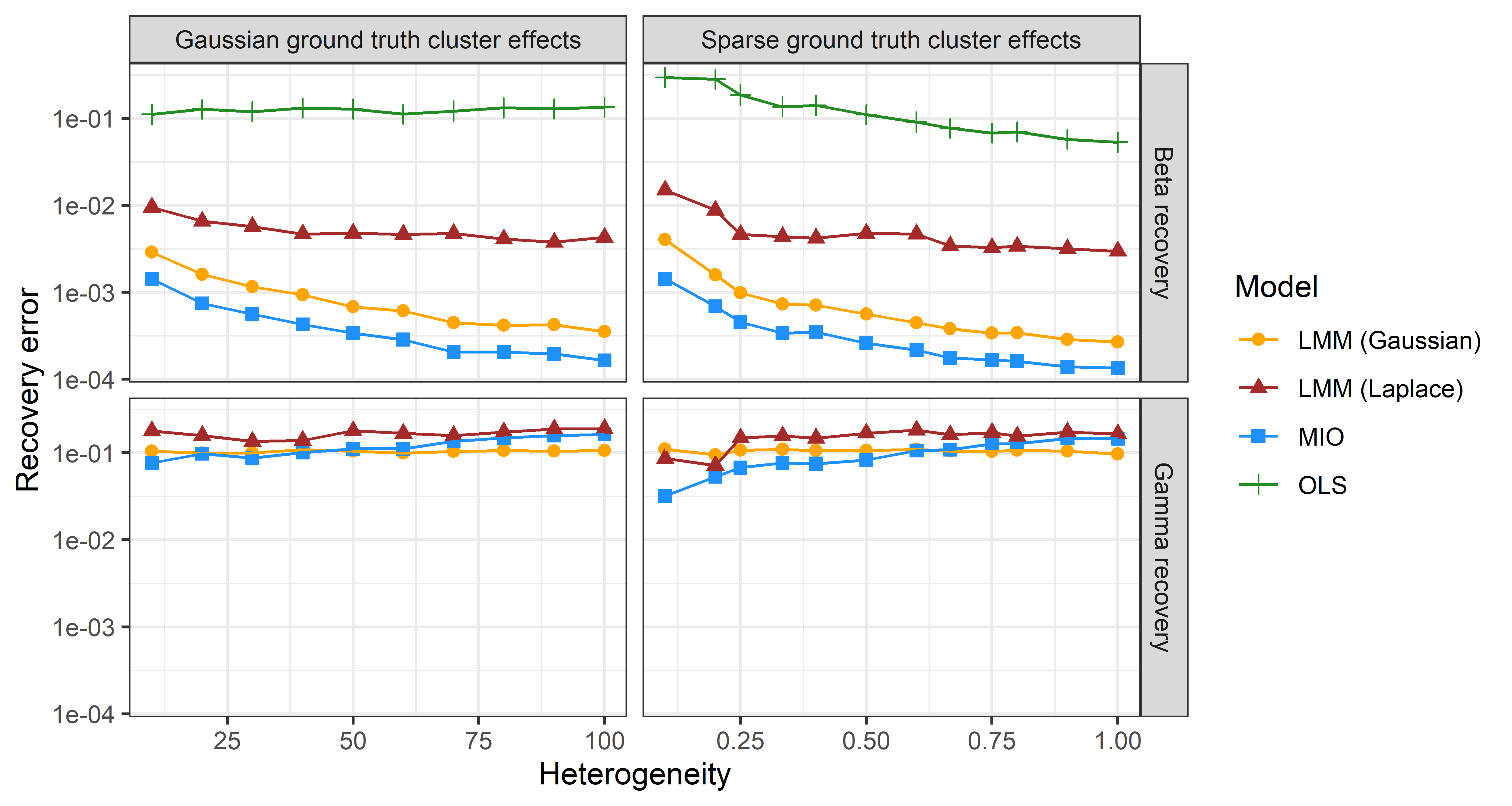}
    \caption{$\ell_2$ causal effect recovery on the log scale ($\beta$ (above) and $\gamma$ (below)) under the simulation scenarios where the cluster-effects are truly Gaussian (left) and truly sparse (right), with a medium dimensional setup (10 clusters, 25 covariates, 50 observations per cluster), where the cluster-effects are truly Gaussian (left) and truly sparse (right)}
    \label{fig:causal_recovery_medium}
\end{figure*}

\begin{figure}[h!]
\centering
    \includegraphics[width = \textwidth]{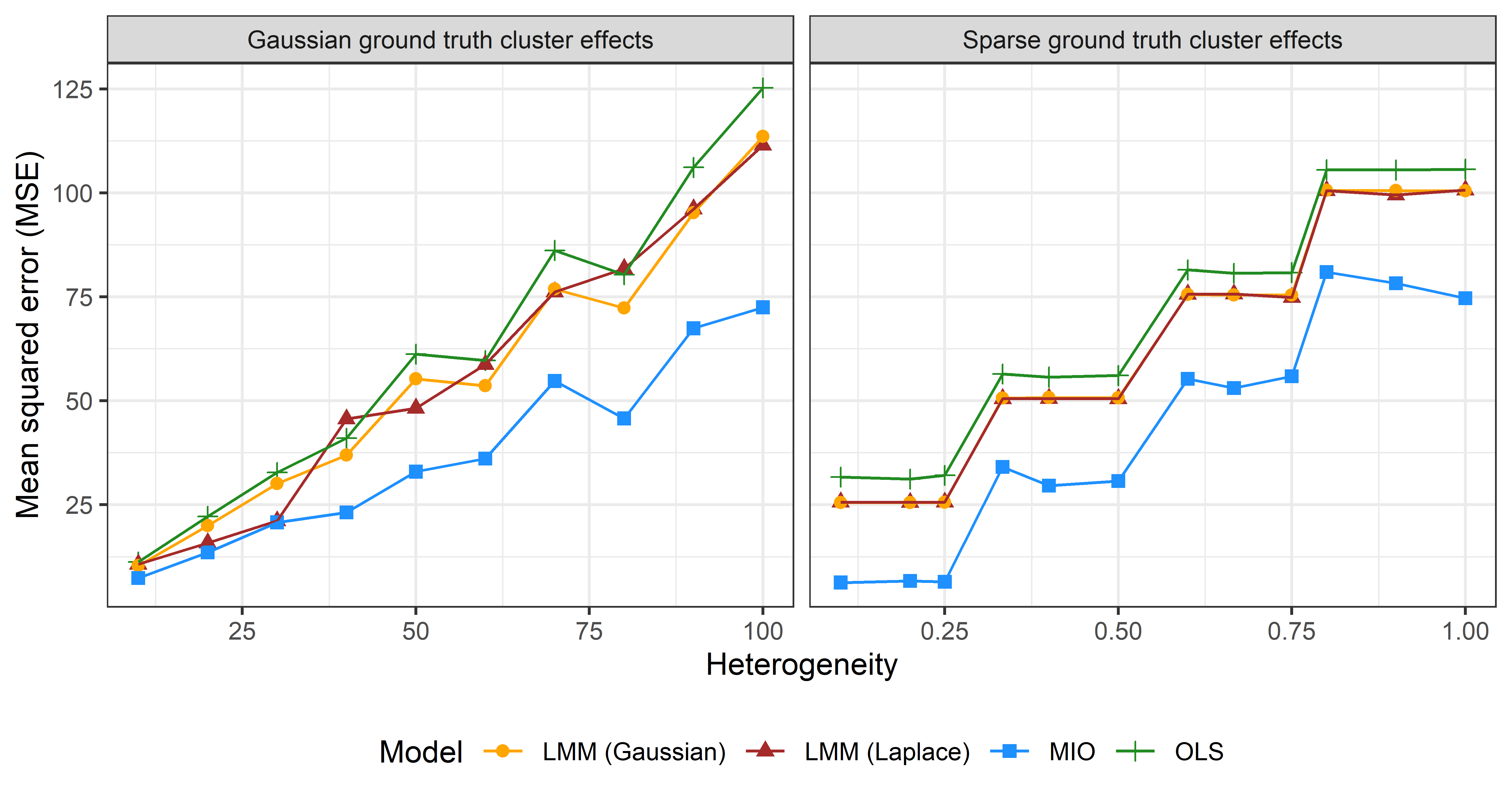}
    \caption{Predictive MSE (on the test set) of the four algorithms in a low dimensional setup (4 clusters, 10 covariates, 50 observations per cluster), where the cluster-effects are truly Gaussian (left) and truly sparse (right) }
    \label{fig:mse_1_low}
\end{figure}

\begin{figure}[h!]
\centering
    \includegraphics[width = \textwidth]{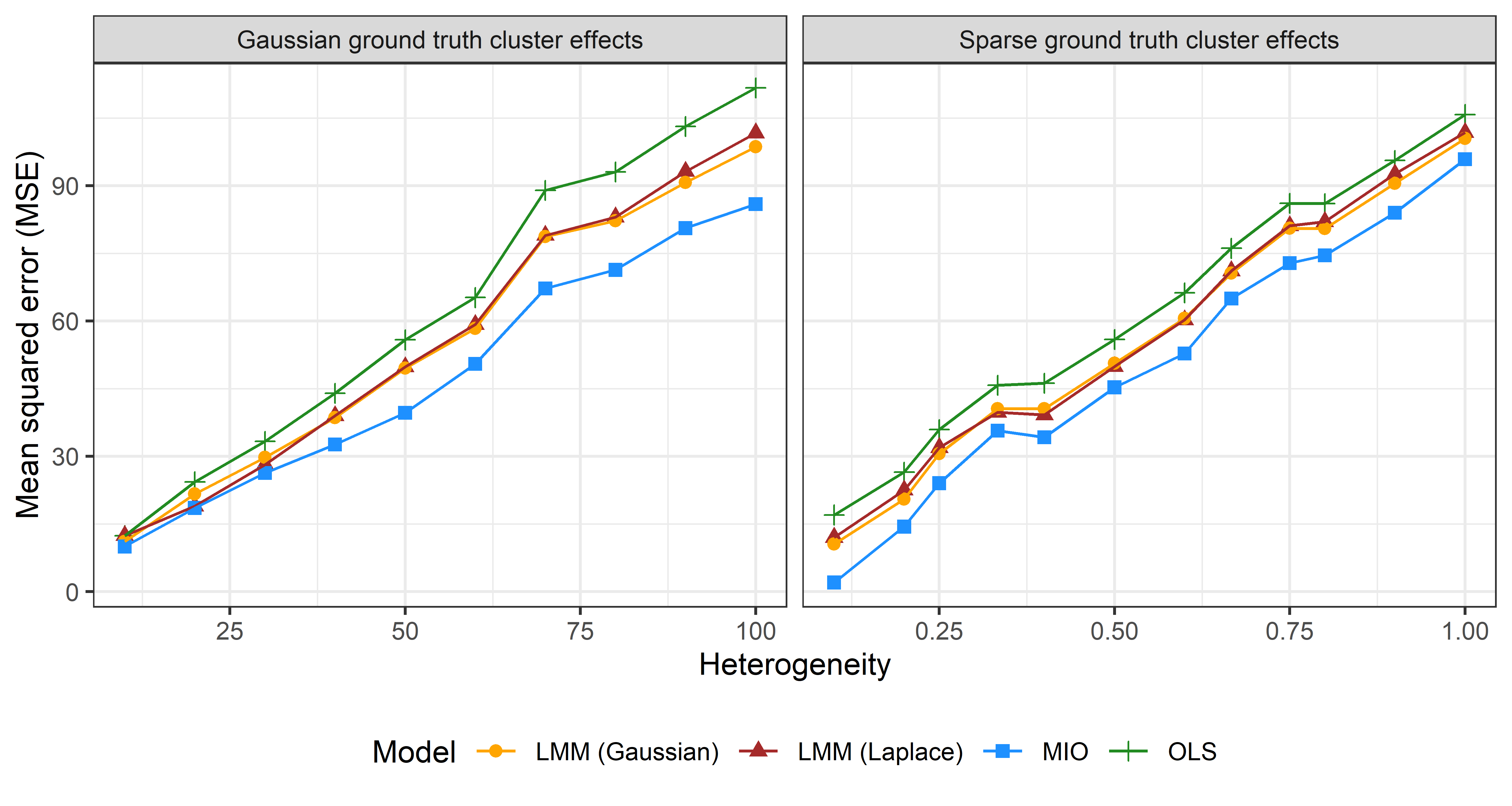}
    \caption{Predictive MSE (on the test set) of the four algorithms in a medium dimensional setup (10 clusters, 25 covariates, 50 observations per cluster), where the cluster-effects are truly Gaussian (left) and truly sparse (right) }
    \label{fig:mse_1_med}
\end{figure}

\clearpage
\renewcommand{\thesection}{C}
\renewcommand{\theHsubsection}{C\thesection}
\section{Data example}
\renewcommand\thetable{\thesection\arabic{table}}  
\setcounter{table}{0}
\renewcommand\thefigure{\thesection\arabic{figure}}  
\setcounter{figure}{0}

\subsection{Protein expression}

We utilize the Mouse Protein Expression Data from the \href{https://archive.ics.uci.edu/ml/datasets/Mice+Protein+Expression}{UCI Machine Learning Repository}. The dataset comprises 77 protein measurements obtained from the brains of 72 mice \citep{Dua:2019, Ahmed_Dhanasekaran_Block_Tong_Costa_Stasko_Gardiner_2015}. Among these mice, 38 are controls and 34 are trisomic, afflicted with Down Syndrome. For each mouse, each protein is measured 15 times. It would be reasonable to postulate a form of cluster structure within the repeated observations for each mouse, irrespective of the different classes delineated in the dataset.

\citet*{Higuera_Gardiner_Cios_2015} probed the proteins that significantly differentiate between classes. Given that the class dependence remains consistent across observations of the same mouse, we can presume a level of cluster effect influencing the expression of these pivotal proteins. Consequently, to assess the efficacy of our algorithm against existing methods, we construct a linear regression problem using these significant proteins to discern if accounting for clustering enhances our predictive capacity.

\citet{Higuera_Gardiner_Cios_2015} identifies 11 proteins as significant in differentiating between at least two classes of mice, inclusive of trisomic and control mice. These 11 proteins, representing a diverse array of biological pathways as per Table 3 of \citet{Higuera_Gardiner_Cios_2015}, will be the focus of our analysis. We aim to examine the effect of other proteins from various biological pathways on these significant proteins.

\begin{figure}[h]
    \centering
    \includegraphics{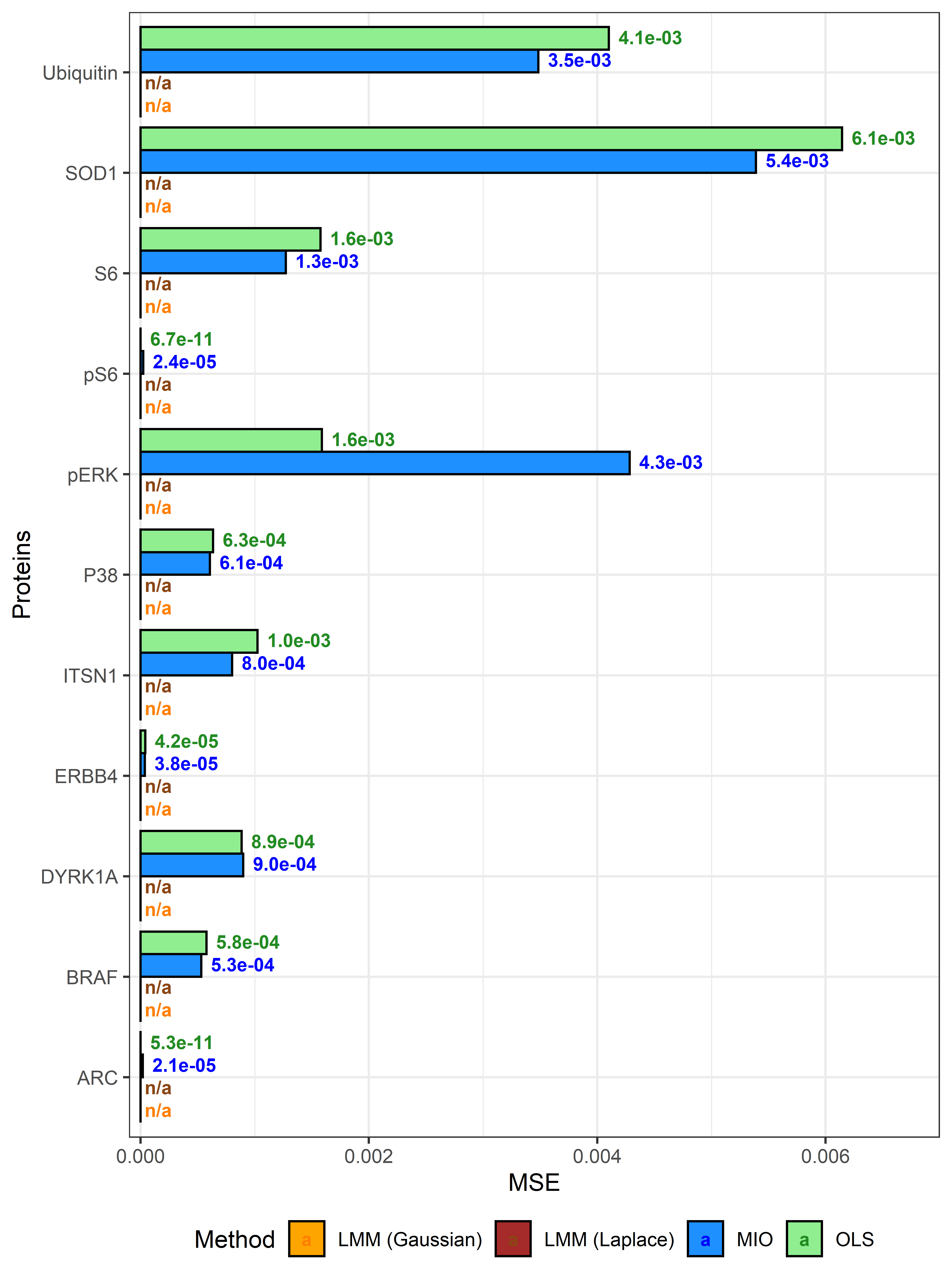}
    \caption{Predictive performance of MIO and OLS on significant proteins from \citet{Higuera_Gardiner_Cios_2015}. \textbf{LMM (Gaussian) and LMM (Laplace) do not produce any results due to convergence issues.}}
    \label{fig:prot}
\vspace{-4mm}
\end{figure}

Employing complete cases ($n=552$), we train a regression model on each of the aforementioned proteins, using the remaining 76 proteins as covariates, and evaluate predictive performance on a reserved test dataset. Figure \ref{fig:prot} displays the predictive performances of the various methodologies.

It is noteworthy that both Gaussian and Laplace LMM approaches fail to converge in this setup, owing to complications in resolving singularities in data that emerge when making distributional assumptions. Despite this, we observe an improvement in predictive performance between Ordinary Least Squares (OLS) regression (no clusters assumed) and our cluster-informed MIO approach for 7 out of the 11 proteins. This suggests the presence of inherent cluster structure in the data that can be exploited for enhanced regression analysis. In this context, LMM is incapable of modeling this clustered nature, illustrating a limitation of distribution-based models; well-behaved data is a prerequisite for models like LMM to function effectively.

\clearpage
\subsection{Student data}

The dataset was obtained from \href{https://archive.ics.uci.edu/ml/datasets/student%2Bperformance}{UCI Machine Learning Repository }. It features information about the academic performance of 382 Portuguese secondary school students, as explored in a study by \citet{cortez2008using}. The dataset encompasses 30 predictors, comprising demographic, social, and school-related factors such as parental education and study time. Continuous predictors were standardized, and dummy variables were introduced for categorical predictors, resulting in a total of 20 predictors. The outcome variables are the students' test scores in Mathematics and Portuguese.

\begin{figure}[h!]
\centering
    \includegraphics[width = 0.8\textwidth]{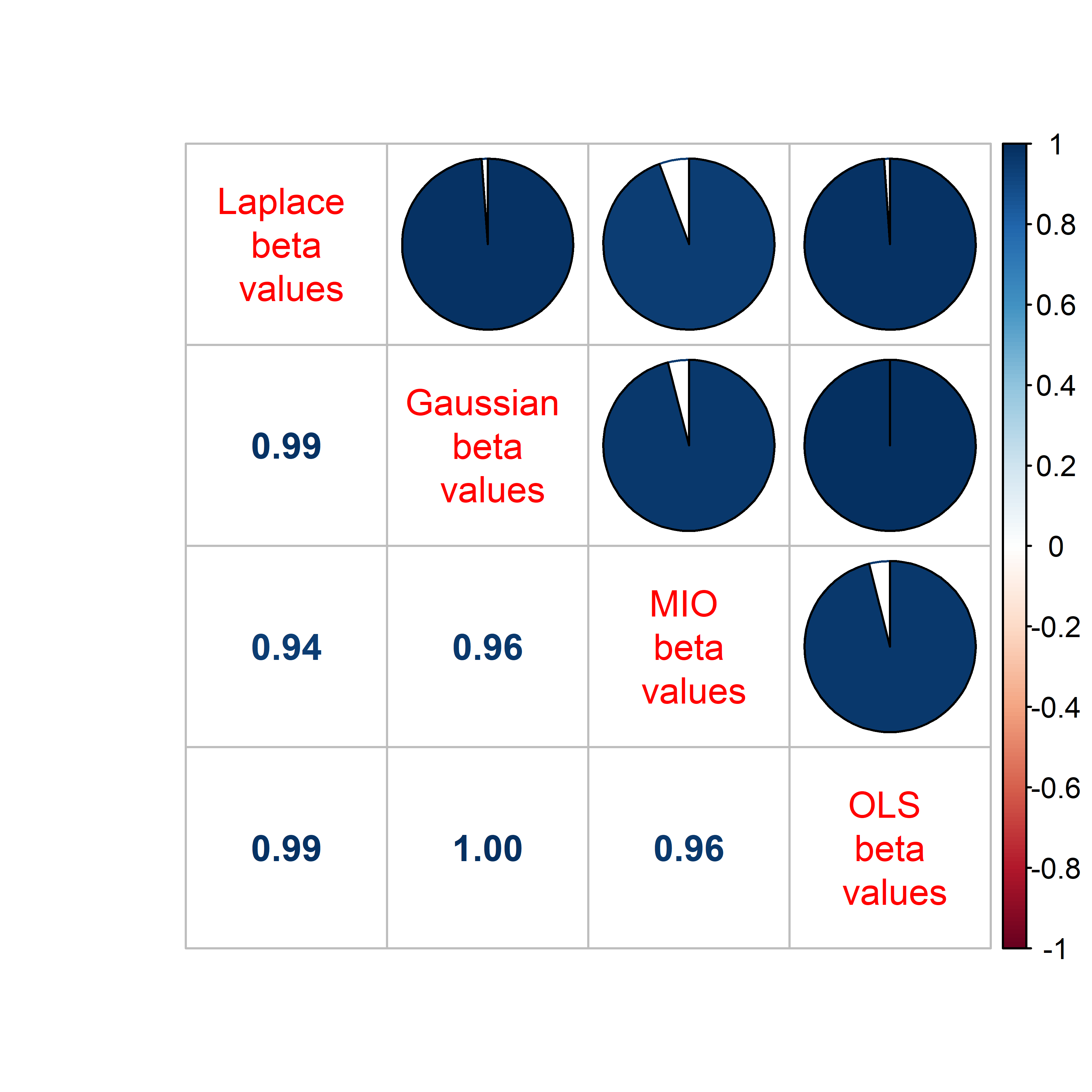}
    \caption{The pie charts on the upper-diagonal portion and values on the lower-diagonal portion represent the correlation of the $\bs{\beta}$ vectors between each pair of methods}
    \label{fig:student_beta_corrs}
\end{figure}

In the main discussion of our paper, we contrasted the cluster-informed rankings yielded by the predictive methods with the rankings derived from raw scores. We observed that the MIO method identifies a distinct set of "exceptional" students compared to the raw score method. To validate that the algorithms exhibit reasonable cluster effects, we examine how well they recover the $\bs{\beta}$ vector. We adopt two strategies for this purpose. The first is to evaluate the population predictive Mean Squared Error (MSE) using the recovered $\bs{\beta}$ vector, and the second is to assess the pairwise concordance of each vector. The benchmark for comparison will be the vector resulting from Ordinary Least Squares (OLS) regression. Table \ref{Tab:bet_recov_student} presents the predictive MSE for each method, which appear to be on a comparable scale. Figure \ref{fig:student_beta_corrs} illustrates the pairwise correlations of the vectors. The evidence suggests that the examination of cluster effects is well warranted as all the algorithms are comparably efficacious in recovering the fixed effects.

\begin{table}[h!]
\caption{The predictive MSE of each regression method. Note that this is a population level prediction, and only leverages $\boldsymbol{\beta}$ values}
\centering
\begin{tabular}{ c|c|c|c|c| } 
\cline{2-5}
 &OLS&LME (Gaussian)&LME (Laplace)&MIO\\
\hline
\multicolumn{1}{|c|}{MSE} & 10.45442& 10.65313& 10.81555& 10.79221\\
\hline
\end{tabular}
\label{Tab:bet_recov_student}
\end{table}

\clearpage
\bibliographystyle{jneurosci}
\bibliography{cemio_bib}